\documentclass[prd,aps,preprintnumbers, showpacs, nofootinbib,notitlepage]{revtex4-1}
\usepackage{epsfig}
\usepackage{amsfonts}
\usepackage{amsmath}
\usepackage{graphicx}
\usepackage{color}
\usepackage{amssymb,amsmath,psfrag,slashed,graphicx}

\begin{document}
	
\preprint{JLAB-THY-19-2907}

\title{Complete Matching for Quasi-distribution Functions in \\ Large Momentum Effective Theory}

\author{Wei Wang$^1$~\footnote{wei.wang@sjtu.edu.cn},
Jian-Hui Zhang$^{2,3}$~\footnote{jianhui.zhang@ur.de},
 Shuai Zhao$^{1,4,5}$~\footnote{szhao@odu.edu}, Ruilin Zhu$^{6}$~\footnote{rlzhu@njnu.edu.cn} }
\affiliation{
$^1$  SKLPPC, MOE KLPPC, School of Physics and Astronomy, Shanghai Jiao Tong University, Shanghai, 200240,   China\\
$^2$Institut f\"ur Theoretische Physik, Universit\"at Regensburg, D-93040 Regensburg, Germany\\
$^3$Center of Advanced Quantum Studies, Department of Physics,
Beijing Normal University, Beijing 100875, China,\\
$^4$Physics Department, Old Dominion University, Norfolk, VA 23529, USA,\\
$^5$Theory Center, Thomas Jefferson National Accelerator Facility, Newport News, VA 23606, USA,\\
$^6$ Department of Physics and Institute of Theoretical Physics,
Nanjing Normal University, Nanjing, Jiangsu 210023, China }

\begin{abstract}
We complete the procedure of extracting parton distribution functions (PDFs) using large momentum effective theory (LaMET) at leading power accuracy in the hadron momentum. We derive a general factorization formula for the quasi PDFs in the presence of mixing, and give the corresponding hard matching kernel at $\mathcal O(\alpha_s)$, both for the unpolarized and for the polarized quark and gluon quasi-PDFs. Our calculation is performed in a regularization-independent momentum subtraction scheme. The results allow us to match the nonperturbatively renormalized quasi-PDFs to normal PDFs in the presence of mixing, and therefore can be used to extract flavor-singlet quark PDFs as well as gluon PDFs from lattice simulations.

\end{abstract}

\maketitle

\section{Introduction}

Understanding the internal structure of hadrons from quarks and gluons --- the fundamental degrees of freedom of QCD Lagrangian --- has been a key goal in hadron physics. However, this is profoundly difficult because it requires solving QCD at large distance scales and thus at strong coupling. In high energy collisions,  the hadron and/or the probe moves nearly at the speed of light, the hadron structure greatly simplifies and can be characterized by certain parton observables such as the parton distribution functions (PDFs), distribution amplitudes (DAs) etc.  The parton observables are defined as the expectation value of lightcone correlations in the hadron state and therefore can not be readily computed on a Euclidean lattice. Currently, the most widely used approach to determine them is to assume a smoothly parametrized form and fit the unknown parameters to a large variety of experimental data (for a recent review, see e.g. Ref.~\cite{Gao:2017yyd}). Lattice efforts on determining the parton observables have been mainly focused on the computation of their moments, which are matrix elements of local operators. The parton observables can be reconstructed in principle if all their moments are known. However, to date only the first few moments can be calculated in lattice QCD~\cite{Martinelli:1987zd,Martinelli:1988xs,Detmold:2001dv,Dolgov:2002zm} due to power divergent mixing between different moments operators and increasing stochastic noise for high moments operators.

In the past few years, a breakthrough has been made to circumvent the above difficulty, which has now been formulated as large momentum effective theory (LaMET)~\cite{Ji:2013dva,Ji:2014gla}. According to LaMET,
a parton observable, instead of its moments, can be directly accessed from lattice QCD using the following procedure: 1) Construct an appropriate static-operator matrix element (quasi-observable) that approaches the parton observable in the infinite momentum limit of the external hadron. The quasi-observable constructed in this way is usually hadron-momentum-dependent but time-independent, and thus can be readily computed on the lattice. 2) Calculate the quasi-observable on the lattice and renormalize it nonperturbatively in an appropriate scheme. 3) Match the renormalized quasi-observable to the parton observable through a factorization formula accurate up to power corrections that are suppressed by the hadron momentum. The existence of such a factorization is ensured by construction; for a proof in the case of isovector quark distribution, see Refs.~\cite{Ma:2014jla,Ma:2017pxb,Izubuchi:2018srq}.

Since LaMET was proposed, much progress has been achieved both in the theoretical understanding of the formalism~\cite{Xiong:2013bka,Ji:2015jwa,Ji:2015qla,Xiong:2015nua,Ji:2014hxa,Monahan:2017hpu,Ji:2018hvs,Stewart:2017tvs,
Constantinou:2017sej,Green:2017xeu,Izubuchi:2018srq,Xiong:2017jtn,Wang:2017qyg,Wang:2017eel,Xu:2018mpf,
Chen:2016utp,Zhang:2017bzy,Ishikawa:2016znu,Chen:2016fxx,Ji:2017oey,
Ishikawa:2017faj,Chen:2017mzz,Alexandrou:2017huk,Constantinou:2017sej,
Green:2017xeu,Chen:2017mzz,Chen:2017mie,Lin:2017ani,Chen:2017lnm,Li:2016amo,
Monahan:2016bvm,Radyushkin:2016hsy,Rossi:2017muf,Carlson:2017gpk,Ji:2017rah,
Briceno:2018lfj,Hobbs:2017xtq,Jia:2017uul,Xu:2018eii,Jia:2018qee,Spanoudes:2018zya,Rossi:2018zkn,Liu:2018uuj,
Ji:2018waw,Bhattacharya:2018zxi,Radyushkin:2018nbf,Zhang:2018diq,Li:2018tpe,Braun:2018brg,Liu:2018tox,
Ebert:2018gzl,Ebert:2019okf,Constantinou:2019vyb,Liu:2019urm,Bhattacharya:2019cme} and in the direct calculation of PDFs from lattice QCD~\cite{Lin:2014zya,Chen:2016utp,Lin:2017ani,Alexandrou:2015rja,Alexandrou:2016jqi,Alexandrou:2017huk,
Chen:2017mzz,Zhang:2017bzy,Chen:2017gck,Alexandrou:2018pbm,Chen:2018xof,Chen:2018fwa,Alexandrou:2018eet,
Lin:2018qky,Fan:2018dxu,Liu:2018hxv,Alexandrou:2019lfo}.
In particular, multiplicative renormalization of both the quark~\cite{Ji:2017oey,
Ishikawa:2017faj,Green:2017xeu} and the gluon~\cite{Zhang:2018diq,Li:2018tpe} quasi-PDF has been established in coordinate space. Nonperturbative renormalization in the regularization-independent momentum subtraction (RI/MOM) scheme as well as a perturbative matching in the same scheme has been carried out for the isovector quark quasi-PDFs in Refs.~\cite{Chen:2017mzz,Stewart:2017tvs,Chen:2018xof,Lin:2018qky} (see also~\cite{Constantinou:2017sej,Alexandrou:2017huk,Alexandrou:2018pbm}). Despite limited volumes and relatively coarse lattice spacings, the state-of-the-art nucleon isovector quark PDFs determined from lattice data at the physical point have shown a reasonable agreement~\cite{Chen:2018xof,Lin:2018qky,Alexandrou:2018pbm}  with phenomenological results extracted from the experimental data~\cite{Dulat:2015mca,Ball:2017nwa,Harland-Lang:2014zoa,Nocera:2014gqa,Ethier:2017zbq}. Of course, a careful study of theoretical uncertainties and lattice artifacts is still needed to fully establish the reliability of the results.

So far the lattice calculations of PDFs have been focused on the isovector quark PDFs only, which do not involve mixing with gluon PDFs and therefore are the easiest to calculate. In the past few years, there has been increasing interest in calculating flavor-singlet quark PDFs and gluon PDFs from lattice QCD. Such calculations are possible only if the renormalization and mixing pattern of gluon quasi-PDFs are fully understood. {The ultraviolet (UV) structure of gluon quasi-PDFs was first studied in Refs.~\cite{Wang:2017qyg,Wang:2017eel} by using a simple cutoff regularization, where it was found that the power divergences cannot be removed by a multiplicative renormalization factor. However, as we pointed out in Ref.~\cite{Zhang:2018diq}, such a cutoff scheme in general breaks gauge invariance in QCD, and therefore obscures the structure of genuine power divergences of the theory. To avoid this, we have chosen in Ref.~\cite{Zhang:2018diq} to work in dimensional regularization and keep track of the power divergences by expanding at $d<4$. For example, at one-loop the linear divergence appears as poles at $d=3$. In this way, we are able to extract the power divergences gauge invariantly. Based on this, we performed a systematic study of the renormalization property of gluon quasi-PDF operators, and showed that with an appropriate choice they are indeed multiplicatively renormalizable.} We have also identified four independent gluon quasi-PDF operators that have an easy implementation on the lattice. Moreover, a general factorization formula for the gluon as well as the quark quasi-PDF in the presence of mixing has been conjectured.

In this paper, we provide all necessary inputs for extracting both the flavor-singlet quark PDF and the gluon PDF from lattice QCD, thereby completing the procedure of calculating PDFs using LaMET at leading power accuracy in the hadron momentum. We explain how to nonperturbatively renormalize the quark and gluon quasi-PDFs, and derive a general factorization formula for the renormalized quasi-PDFs in the presence of mixing, following the operator product expansion (OPE) method in Refs.~\cite{Ma:2017pxb,Izubuchi:2018srq}. We then present the complete one-loop results for the hard matching kernels that appear in the factorization of quasi-PDFs. {The computation of the matching kernel has been considered in Ref.~\cite{Wang:2017qyg}, but in a scheme that is inappropriate for lattice implementation.}

The rest of the paper is organized as follows: In Sec.~\ref{sec:renormalization_factorization}, we briefly review the renormalization and factorization of quark and gluon quasi-PDFs. In Sec.~\ref{sec:1loopmatching_unpolarized}, we present our one-loop calculation of the hard matching kernel connecting the RI/MOM renormalized quasi-PDFs to the PDFs in $\overline{\rm MS}$ scheme, with a particular focus on the unpolarized case. Sec.~\ref{sec:1loopmatching_polarized} is devoted to the polarized case. We then conclude in Sec.~\ref{conclusion} and give some computational details in the Appendix.

\section{Renormalization and factorization of quark and gluon quasi-PDFs \label{sec:renormalization_factorization}}

In this section, we give a brief review of the renormalization and factorization of quark and gluon quasi-PDFs in LaMET.

\subsection{Quasi-PDFs in LaMET}
\label{qpdfsdef}

In high-energy collisions, the PDFs are defined as the hadron matrix elements of quark and gluon nonlocal correlators along the lightcone. For example, the unpolarized quark distribution is defined as
\begin{align}\label{eq:pdf}
f_{q_i/H}(x,\mu)=\!\int\!\! \frac{d\xi^-}{4\pi} \, e^{-ixP^+\xi^-}
\! \big\langle P \big| \bar{q}_i(\xi^-) \gamma^+
W(\xi^-\!,0)
q_i(0) \big|P\big\rangle
\end{align}
for a given flavor $i$, where  $x= k^+/P^+$ is the longitudinal momentum fraction carried by the quark of flavor $i$.  $\mu$ is the renormalization scale in the $\overline{\text{MS}}$ scheme,  $P^\mu=(P^0,0,0,P^z)$ is the hadron momentum, $\xi^\pm = (t\pm z)/\sqrt{2}$ are the lightcone coordinates, and
\begin{align}
W(\xi^-,0) &= \exp\bigg(-ig \int_0^{\xi^-}d\eta^- A^+(\eta^-) \bigg) \,
\end{align}
is the Wilson line inserted to maintain the gauge invariance of the nonlocal correlator. The $A^+=A^+_a t^a$ with $t^a$ being the generators in the fundamental representation of color $SU(3)$ group.

Analogously, the unpolarized gluon distribution can be defined as~\cite{Collins:2011zzd}
\begin{eqnarray}
    f_{g/H}(x,\mu) = \int \frac{d\xi^-}{2\pi x P^+} e^{-ixP^+\xi^-}  \langle P|F^{+i}_a(\xi^-)
  {\cal W}(\xi^-,0)F^{+i}_a(0) |P\rangle,\label{eq:gluon_light-cone-PDF}
\end{eqnarray}
where $F^{\mu\nu}_a= \partial^\mu A^{\nu}_a - \partial^\nu A^{\mu}_a - gf_{abc} A^{\mu}_b A^{\nu}_c$ is the gluon field strength, and $i$ runs over the transverse indices. The above Wilson line $\cal W$  takes a similar form as the quark case, but is defined in the adjoint representation.

The quark and gluon PDFs defined above can not be directly  computed on the lattice due to their real-time dependence. However, according to LaMET, they can be extracted from lattice calculations of appropriately constructed quasi-PDFs via a factorization procedure. For the unpolarized quark PDF, a well-suited quasi-PDF candidate is given by
\begin{eqnarray}\label{qlcDef}
{ \tilde  f}_{q_i/H}(x,\mu, P^z) &=& N\int \frac{dz}{4\pi} e^{iz  xP^z}  \langle P|\overline{q}_i (z)
   \Gamma  W(z, 0) q_i(0) |P\rangle,
\end{eqnarray}
where $z$ is a spatial direction, $\Gamma=\{\gamma^z, \gamma^t\}$ is a Dirac matrix with the corresponding normalization factor $N=\{1, P^z/P^t\}$, respectively. As shown in Ref.~\cite{Ji:2017oey}, the renormalization of the quark quasi-PDF defined above is of a multiplicative form so that the matrix elements at different $z$ do not mix with each other. In addition, the choice with $\Gamma=\gamma^t$ has the advantage of avoiding mixing with the scalar PDF when a non-chiral lattice fermion is used~\cite{Constantinou:2017sej,Chen:2017mie}. We will focus on this choice in the rest of the paper.

In comparison with the quark case, what is the most appropriate operator to define the gluon quasi-PDF is less obvious. In principle, one can use
\begin{eqnarray}
O_g^{\mu\nu}(z,0)= F^{\mu \alpha}(z){\cal W}(z,0)F_{\alpha}^{\;\nu}(0), \label{eq:generic_gluon}
\end{eqnarray}
with $\mu, \nu=\{t,z\}$ and $\alpha$ running either over all Lorentz indices or only over transverse indices.
However, such a choice could in principle  mix with other relevant  operators under renormalization. Using the auxiliary field approach~\cite{Dorn:1986dt}, we have explicitly shown~\cite{Zhang:2018diq} that different components of $O^{\mu\nu}$ indeed renormalize differently, which complicates the construction of appropriate gluon quasi-PDFs. A brief review of the formalism used in Refs.~\cite{Zhang:2018diq} and~\cite{Ji:2017oey} will be given in the forthcoming subsections.
Nevertheless, we have identified four gluon operators~\cite{Zhang:2018diq} that are multiplicatively renormalizable and therefore are suitable for defining the gluon quasi-PDF. These operators are
\begin{align}\label{gluonqpdfop}
  O_g^{(1)}(z, 0)&\equiv    F^{ti}(z){\cal W}(z,0)F_{i}^{\;t}(0), \;\;\;
  O_g^{(2)}(z, 0)\equiv F^{zi}(z){\cal W}(z,0)F_i^{\;z}(0), \nonumber\\
  O_g^{(3)}(z, 0)&\equiv F^{ti}(z){\cal W}(z,0)F_i^{\;z}(0), \;\;\;
  O_g^{(4)}(z, 0)\equiv F^{z\mu}(z){\cal W}(z,0)F_{\mu}^{\; z}(0),
\end{align}
where a summation over transverse (all) components is implied for $i\, (\mu)$. The corresponding gluon quasi-PDF is then defined as
\begin{eqnarray}\label{gpdfdef}
  \tilde f^{(n)}_{g/H}(x,\mu, P^z) &=& N^{(n)} \int \frac{dz}{2\pi  x  P^z} e^{iz x P^z}  \langle P|O_g^{(n)}(z, 0) |P\rangle.
\end{eqnarray}
The normalization factors are chosen  by
\begin{eqnarray}
 N^{(2)}=N^{(4)}=1, \;\;\; N^{(1)}= \frac{(P^z)^2}{(P^t)^2},\;\;\;N^{(3)}= \frac{P^z}{P^t},
\end{eqnarray}
so that all partonic gluon PDFs at tree-level are
\begin{eqnarray}\label{ggtree}
 \tilde f^{(n,0)}_{g/g}(x,\mu, P^z) = \delta(x-1),
\end{eqnarray}
with the hadron state $H$ being replaced by a gluon state. Note that in the above result (also in the sections below unless stated otherwise) we have ignored the contributions from the crossed diagrams, which correspond to interchanging the contraction between the two external gluons and gluon fields from the operators $O_g^{(n)}$. These crossed diagrams contribute to $x<0$ and can be easily obtained from $\tilde f^{(n)}_{g/H}(x)=-\tilde f^{(n)}_{g/H}(-x)$.

All above gluon quasi-PDF operators are defined in terms of an adjoint gauge link. Alternatively, these operators  can also be parametrized  using gauge links in the fundamental representation $U(z_2,z_1)$~\cite{Dorn:1981wa,Polyakov:1980ca,Gervais:1979fv,Arefeva:1980zd,Dorn:1980hs,Dorn:1986dt}. Taking  the operator $O^{(3)}_g$ as an example, one could  use
\begin{eqnarray}\label{fundrep}
O_g^{(3)}(z_2, z_1)=2\,{\rm Tr}[F^{ti}(z_2)U(z_2, z_1)F_{i}^{\;z} (z_1)U(z_1, z_2)].
\end{eqnarray}
Here $F^{\mu\nu}=F^{a}_{\mu\nu}t^a$ and $t^a$ is the  generator in the fundamental representation with ${\rm tr}[t^at^b]=1/2\delta^{ab}$. We  stress that Eq.~(\ref{eq:generic_gluon}) facilitates the renormalization study of gluon quasi-PDFs, whereas Eq.~(\ref{fundrep}) makes the implementation on the lattice simpler.  In the following, we will mainly focus on the definition Eq.~(\ref{eq:generic_gluon}), as has been done in Ref.~\cite{Zhang:2018diq}, but the results also apply to Eq.~({\ref{fundrep}}).

In the forthcoming subsections, we briefly review the renormalization of quasi-PDFs in the auxiliary field approach, following our earlier work in Refs.~\cite{Ji:2017oey,Zhang:2018diq}. Other studies have been available using a similar formalism~\cite{Green:2017xeu} or using the Feynman diagrammatic approach~\cite{Ishikawa:2017faj,Li:2018tpe}.

\subsection{Auxiliary Field Approach}

In the auxiliary field approach~\cite{Dorn:1986dt}, one introduces an auxiliary ``heavy quark'' field into the QCD interaction  such that the Wilson line can be reinterpreted as a two-point function of the auxiliary field. For the quark/gluon quasi-PDF, this auxiliary field is chosen to be in the fundamental/adjoint representation of color $SU(3)$ group, respectively. Similar with the  ordinary heavy quark, the auxiliary ``heavy quark'' has trivial spin degrees of freedom. An advantage of this approach is to convert the study of   renormalization of nonlocal operators into the analysis of two local operators.   In the following we present, as an example, the auxiliary Lagrangian that can be used to study quark quasi-PDFs, while for gluon quasi-PDFs the procedure is completely analogous.

The effective Lagrangian with  an auxiliary fundamental ``heavy quark'' field (denoted as $Q$) can be written as
\begin{eqnarray}\label{efflag}
\mathcal L=\mathcal L_{\rm{QCD}}+\overline{Q}(x)i n\cdot DQ(x),
\end{eqnarray}
where $D_\mu=\partial_\mu+igt_a A_{a,\mu}$ is the covariant derivative in the fundamental representation. The unit vector $n^\mu$ is chosen as  $n^\mu=(0,0,0,1)$.

As shown in Ref.~\cite{Ji:2017oey}, the two-point function of the auxiliary ``heavy quark'' is evaluated as
\begin{eqnarray}
\int {\cal D}\overline {Q}{\cal D}{Q}\, {Q}(x)\overline {Q}(y) e^{i\int d^4x {\cal L}}=S_{Q}(x, y) e^{i\int d^4x {\cal L_{\text{QCD}}}}.
\end{eqnarray}
The above equation holds  up to a determinant ${\rm det}(in\cdot D)$ which  can be absorbed into the normalization of the generating functional~\cite{Mannel:1991mc}. The   propagator $S_{Q}(x, y)$ in the above   is the Green function of $n\cdot D$ operator with
\begin{eqnarray}\label{2pointfunceq}
n\cdot D\, S_{Q}(x, y)=\delta^{(4)}(x-y),
\end{eqnarray}
The  solution
\begin{align}\label{wilsonlinesol}
S_{Q}(x,y)
&=\theta(x^z-y^z)\delta(x^0-y^0)\delta^{(2)}(\vec x_\perp-\vec y_\perp)W(x^z,y^z),
\end{align}
can be derived with  an appropriate boundary condition. One should notice that Eq.~(\ref{wilsonlinesol}) is nothing but a spacelike Wilson line along the $z$ direction. One can always restrict oneself to $x^z>y^z$, without   loss of generality.

\subsection{Renormalization of Quasi-PDFs in Auxiliary Field Approach}

\subsubsection{Quark Quasi-PDFs}

From the discussions above, one can see that  the Wilson line $W(z_2,z_1)$ appearing in the quark quasi-PDFs can be replaced by the product of two auxiliary ``heavy quark'' fields ${Q}(z_2)\overline {Q}(z_1)$.
The quark bilocal operator
\begin{eqnarray}
O_{q_i}(z_2,z_1) = \bar q_i (z_2) \Gamma W(z_2,z_1)  q_i(z_1)
\end{eqnarray}
then reduces to the product of two local composite operators
\begin{eqnarray}
{\cal O}_{q_i}(z_2,z_1) = \bar q_i(z_2) \Gamma  {Q}(z_2)   {\overline Q}(z_1)q_i(z_1) \equiv \bar j(z_2) j(z_1),
\end{eqnarray}
with
\begin{eqnarray}
\bar j(z_2)= \bar q_i(z_2)\Gamma {Q}(z_2), \;\;\; j(z_1) = {\overline Q}(z_1) q_i(z_1).
\end{eqnarray}
Since the ``heavy quark'' has trivial spin degrees of freedom, one can also move the Dirac matrix $\Gamma$ into $j(z_1)$.

In dimensional regularization (DR), the local operators $\bar j(z_2), j(z_1)$ are ``heavy-to-light'' like and  are multiplicatively renormalized:
\begin{eqnarray}
 \bar j(z_2)= Z_{\bar j} \bar j_R(z_2), \;\;\;  j(z_1)= Z_{  j}   j_R(z_1),
\end{eqnarray}
with ($D=4-2\epsilon$)
\begin{eqnarray}
Z_{\bar j}= Z_j= 1+\frac{\alpha_s}{2\pi\epsilon}+\mathcal O(\alpha_s^2).
\end{eqnarray}
When the auxiliary field is integrated out, the nonlocal operator renormalizes as~\cite{Ji:2017oey}
\begin{eqnarray}
O_{q_i,R}(z_2,z_1)= Z_{\bar j}^{-1}Z_{  j}^{-1} \bar q_i(z_2) \Gamma W(z_2,z_1) q_i(z_1).
\end{eqnarray}

In lattice regularization, when going beyond leading-order perturbation theory, the self-energy of the heavy quark generates a linear divergence that does not show up in DR. Such a linear divergence can be absorbed into an effective mass counterterm,
\begin{equation}\label{effmass}
    \delta {\cal L}_m =  -\delta m\overline{Q} Q,
\end{equation}
where $\delta m\sim{\cal O}(1/a)$ with $a$ being the lattice spacing~\cite{Maiani:1991az}. As shown in Ref.~\cite{Ji:2017oey}, apart from the structures given in the Lagrangian Eq.~(\ref{efflag}), this is the only possible renormalizable counterterm allowed by the symmetry of the theory. Moreover, Becchi-Rouet-Stora-Tyutin (BRST) invariance requires a dependence of $\delta m$ on the signature of $n$ in Eq.~(\ref{efflag})~\cite{Dorn:1986dt}, For a spacelike $n^\mu$, $\delta m=i\delta {\bar m}$ is imaginary.

Including the effective mass term Eq.~(\ref{effmass}) into the Lagrangian and integrating out the auxiliary ``heavy quark'', we obtain the following renormalization for the nonlocal quark bilinear operator~\cite{Ji:2017oey}
\begin{eqnarray}
O_{q_i,R}(z_2,z_1)= Z_{\bar j}^{-1}Z_{  j}^{-1}e^{\delta \bar m|z_2-z_1|}   \bar q_i(z_2) \Gamma W(z_2,z_1) q_i(z_1).
\end{eqnarray}

\subsubsection{Gluon Quasi-PDFs}

For the nonlocal gluon quasi-PDF operators, the desired auxiliary Lagrangian has exactly the same form as that for the quark, except that now the auxiliary ``heavy quark'' and the covariant derivative are defined in the adjoint representation. To distinguish from the fundamental auxiliary field  used in the previous subsection, we denote the adjoint field  as $\mathcal Q$ below.

With the auxiliary $\mathcal Q$, one  can decompose  the nonlocal gluon operator in Eq.~(\ref{gluonqpdfop}) into  the product of two local composite operators. For example, the $O_g^{(3)}$ has:
\begin{eqnarray}\label{localop}
 \mathcal O_g^{(3)}(z_2, z_1) = J_1^{ti}(z_2) {\overline J}_{1, i}^{\;\;\;z}(z_1),
\end{eqnarray}
where
\begin{eqnarray}
J_1^{ti}(z_2)=F^{ti}_a(z_2)\mathcal Q_a(z_2),\, {\overline J}_{1, i}^{\;\;\;z}(z_1)=\overline{\mathcal Q}_b(z_1)F_{b, i}^{\;\;\;z}(z_1).
\end{eqnarray}
Again the renormalization of the gluon quasi-PDF operator then reduces to the renormalization of the local gluon composite operators $J_1, {\overline J}_1$, which is easier to handle.

The operator  $J_1^{\mu\nu}$ can  mix with operators of the same or lower mass dimension under renormalization.  The mixing operators can be of the following three types: 1) gauge-invariant operators, 2) BRST exact operators or operators that are the BRST variation of some other operators, 3) operators that vanish by equation of motion~(see e.g. \cite{Collins:1984xc}).
Let us start with the renormalization in DR for simplicity.
In DR, it has been shown in Refs.~\cite{Dorn:1981wa,Dorn:1986dt,Zhang:2018diq} that, the   operators that   mix with $J_1^{\mu\nu}$ are
\begin{align}\label{mixingop}
J_2^{\mu\nu}&=n_{\rho}(F^{\mu\rho}_a n^\nu-F^{\nu\rho}_a n^{\mu}){\mathcal Q}_a/n^2,\nonumber\\
J_3^{\mu\nu}&=(-in^\mu A_a^{\nu}+in^\nu A_a^\mu)((in\cdot D-\delta m) \mathcal Q)_a/n^2,
\end{align}
where a potential mass term for the auxiliary field is included. Such a mass term is absent in DR, but can be generated by radiative corrections in a cutoff regularization such as the lattice regularization.  The operator $J_2^{\mu\nu}$ is gauge invariant, whereas $J_3^{\mu\nu}$ is proportional to the massive equation of motion of $\mathcal Q$ and therefore vanishes in a physical matrix element. The above mixing pattern has been verified by us in an explicit one-loop calculation~\cite{Zhang:2018diq}.

The renormalization of the above three types of composite operators then takes  the following form
\begin{eqnarray}\label{oprenormmix}
\begin{pmatrix}
 J_{1, R}^{\mu\nu} \\  J_{2, R}^{\mu\nu} \\ J_{3, R}^{\mu\nu}
\end{pmatrix}
&=&
\begin{pmatrix}
Z_{11} & Z_{12} & Z_{13}\\ 0 & Z_{22} & Z_{23} \\ 0 & 0 & Z_{33}
\end{pmatrix}
\begin{pmatrix}
 J_1^{\mu\nu}\\  J_2^{\mu\nu} \\ J_3^{\mu\nu}
\end{pmatrix}
,
\end{eqnarray}
where the  mixing matrix $Z$ is triangular. However
the renormalization constants in Eq.~(\ref{oprenormmix}) are not all independent, as demonstrated in Ref.~\cite{Zhang:2018diq}. A first observation is the degeneracy of  $J_2^{z\mu}$ and  $J_1^{z\mu}$, which leads to the following relation between the renormalization constants in Eq.~(\ref{oprenormmix})
\begin{eqnarray}\label{Zrelation}
Z_{11}+Z_{12}=Z_{22}, \;\;\;  Z_{13}=Z_{23}.
\end{eqnarray}
An explicit one-loop calculation in Ref.~\cite{Dorn:1981wa} has indeed verified the above expectation.
Since  $J_2^{\mu z}$ is not independent, it  can be ignored in the studies of operator renormalization.
In addition,
Eqs.~(\ref{oprenormmix}) and (\ref{Zrelation}) indicate that $J_1^{z\mu}$ and $J_1^{ti}(i=1,2)$ renormalize independently. As a result, the renormalization pattern can be simplified to
\begin{eqnarray}\label{oprenormmixsim2x2}
\begin{pmatrix}
 J_{1, R}^{z\mu} \\ J_{3, R}^{z\mu}
\end{pmatrix}
&=&
\begin{pmatrix}
Z_{22} & Z_{13} \\  0 & Z_{33}
\end{pmatrix}
\begin{pmatrix}
 J_1^{z\mu} \\ J_3^{z\mu}
\end{pmatrix},\hspace{2em}
J_{1, R}^{ti}=Z_{11} J_1^{ti}, \;\;\; J_{1, R}^{ij}=Z_{11} J_1^{ij}.
\end{eqnarray}
The reason that
($J_{1}^{ti}$, $J_{1}^{ij}$) and $J_1^{z\mu}$ have different renormalizations is due to the Lorentz symmetry breaking  in the presence of a four-vector $n^\mu$ along the $z$ direction.

To extract the UV divergences, in particular the genuine power divergences inherited from the operator $J_1^{\mu\nu}$, one should  introduce a proper UV regulator in a gauge-invariantly manner. In Ref.~\cite{Zhang:2018diq}, we worked in DR and kept track of the linear divergences by expanding the results around $d=3$, as the linear divergences appear as poles around $d=3$ at one-loop.

\begin{figure}[tbp]
\includegraphics[width=.15\textwidth]{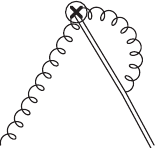}
\hspace*{5em}
\includegraphics[width=.15\textwidth]{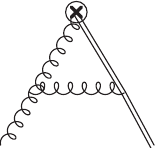}
\caption{One-loop corrections  with  linear divergences to  the $J_1^{\mu\nu}$. The double line represents the auxiliary adjoint  field $\mathcal Q$.}
\label{1looplindivdiag}
\end{figure}

The one-loop diagrams that give rise to linearly divergent contributions to the operator $J_1^{\mu\nu}$ are shown in Fig.~\ref{1looplindivdiag}, and other diagrams are neglected. We have performed a detailed  calculation in coordinate space in Ref.~\cite{Zhang:2018diq}, and the result reads
\begin{align}
I_1^{\rho\nu}&= \frac{\alpha_s C_A}{\pi}\Big\{\frac{1}{d-4}(A_a^\nu n^\rho-A_a^\rho n^\nu) n\cdot \partial \mathcal Q_a/n^2+\frac{\pi\mu}{d-3}\big(n^\rho A_a^\nu-n^\nu A_a^\rho\big)\mathcal Q_a+reg.\Big\},\nonumber\\
I_2^{\rho\nu}&=\frac{\alpha_sC_A}{\pi}\Big\{\frac{1}{d-4}\Big[\frac{1}{4}F_a^{\rho\nu}{\mathcal Q}_a+\frac{1}{2}\big(F_a^{\rho\sigma}n^\nu n_\sigma-F_a^{\nu\sigma}n^\rho n_\sigma\big)\mathcal Q_a/n^2+\frac 1 2 (A_a^\rho n^\nu-A_a^\nu n^\rho) n\cdot \partial \mathcal Q_a/n^2\Big]\nonumber\\
&-\frac{\pi\mu}{d-3}\big(n^\rho A_a^\nu-n^\nu A_a^\rho\big)\mathcal Q_a+reg.\Big\}, \label{eq:pole_auxiliary}
\end{align}
where $\mu$ is the regularization scale and $reg.$ denotes regular terms at both $d=4$ and $d=3$. Combining the results in Eq.~\eqref{eq:pole_auxiliary}, we find  the linear divergences cancel. Our results show an identical mixing pattern as in Ref.~\cite{Dorn:1981wa} (note the difference in the normalization of the direction vector).

Based on the renormalization analysis above, one can derive useful building blocks for the construction of appropriate gluon quasi-PDFs.
To this end, we may use  one of the indices in $J_1^{\mu\nu}$ with $z$ or $t$ and let the other indices run either over all Lorentz components or over the transverse components only. It is necessary to point out that the  operator $J_3^{\mu\nu}$ only yields contact terms when integrating out the ``heavy quark'' field, since the equation of motion operator acting on the ``heavy quark'' propagator yields a $\delta$-function. The nonvanishing contact terms  at $z_2=z_1$ indicate that an extra renormalization is required when the distance between two local composite operators shrinks to zero. When $z_1\neq z_2$, the operator $J_3^{\mu\nu}$ is irrelevant and can be ignored.

In a cutoff scheme like the lattice regularization, the mass term of the $\mathcal Q$  could   appear beyond leading order in perturbation theory even if it does not exist at leading order.
This is indeed what happens here. In perturbation theory, $m=\delta m$ starts from $O(\alpha_s)$.
Such a mass term serves the purpose to absorb  power divergences arising from the Wilson line self-energy.  Apart from this, there is no other power divergence in the theory.  Therefore in a gauge-invariant cutoff scheme, the operator renormalization remains the same as Eq.~(\ref{oprenormmixsim2x2}).

With $J_{1,R}^{zi}$, $J_{1,R}^{ti}$, $J_{1,R}^{z\mu}$, and their conjugate as the building blocks, four multiplicatively renormalizable unpolarized gluon quasi-PDF operators have been constructed~\cite{Zhang:2018diq}, and their explicit form has been given in Sec.~\ref{qpdfsdef}. To illustrate how the gluon quasi-PDF operators renormalize, let us take
\begin{eqnarray}
 {\cal O}^{(3)}_{g,R}(z_2,z_1) \equiv  J_{1,R}^{ti}(z_2) \overline J_{1,R,i}^{\;\;\;\;\;\; \; z}(z_1)
\end{eqnarray}
as an example. When the auxiliary ``heavy quark'' field is integrated out, the  $ {\cal O}^{(3)}_{g,R}(z_2,z_1)$ operator renormalizes multiplicatively as ($\delta m=i\overline{\delta { m}}$)
\begin{eqnarray}\label{O3renorm}
{ {O}^{(3)}_{g,R}(z_2, z_1) = (F^{ti}(z_2) {\cal W}(z_2,z_1) F_{i}^{\;\; z}(z_1))_R=Z_{11}Z_{22} e^{\overline{\delta { m}}|z_2-z_1|}F^{ti}(z_2) {\cal W}(z_2,z_1) F_{i}^{\;z}(z_1)}.
\end{eqnarray}
The renormalization of other operators is  analogous  with different renormalization factors~\cite{Zhang:2018diq}.

Actually, the operators  ${\cal O}^{(i)}_{g,R}(i=1,2,3,4)$ belong to the same universality class~\cite{Hatta:2013gta} and  differ  only by power corrections in the large momentum limit.   Bearing in mind the different renormalizations of ${\cal O}^{(i)}_{g,R}$, one may use any combination of them to study the gluon quasi-PDF.   A notable example is
\begin{eqnarray}
 {  O}^{(5)}_{g,R}(z_2,z_1) \equiv   (F^{t\mu}(z_2) {\cal W}(z_2,z_1) F^{t}_{\;\; \mu}(z_1))_R =    -  {  O}^{(1)}_{g,R}(z_2,z_1)-  {  O}^{(2)}_{g,R}(z_2, z_1)-{  O}^{(4)}_{g,R}(z_2, z_1).
\end{eqnarray}
This operator (minus the trace term) has been used in a recent simulation~\cite{Fan:2018dxu}. Since the renormalizations for ${  O}^{(1)}_{g,R}(z_2,z_1)$ and  $ {  O}_{g,R}^{(2,4)}(z_2, z_1)$ are different, ${  O}_{g,R}^{(5)}(z_2,z_1)$ is not multiplicatively renormalizable.

\subsection{Renormalization in RI/MOM Scheme and Implementation on Lattice}

From the discussions above, it is clear that the nonlocal operators at different $z$ do not mix under renormalization. This allows us to carry out a nonperturbative renormalization of the quasi-PDF in the following manner: 1) Calculate the endpoint renormalization factors (e.g. $Z_{\{11, 22\}}$ in Eq.~(\ref{O3renorm})) and the Wilson line mass counterterm ($\overline{\delta m}$ in Eq.~(\ref{O3renorm})) nonperturbatively. The calculation of the former is rather straightforward, while the latter can be determined by using the static-quark potential for the renormalization
of Wilson loops~\cite{Musch:2010ka}. This has been used in early studies of nucleon PDFs and meson DAs~\cite{Chen:2016fxx,Zhang:2017bzy,Chen:2017gck}. 2) Calculate the renormalization factors as a whole for each $z$. This is analogous to the renormalization of local composite operators, which is usually carried out in the RI/MOM scheme~\cite{Martinelli:1994ty} on the lattice. In the RI/MOM scheme, the renormalization of local composite operators is done by demanding that the counterterm cancels all loop contributions to their matrix element between off-shell external states at specific momenta~\cite{Stewart:2017tvs,Chen:2017mzz} (for the application to quark and glue momentum fractions see Ref.~\cite{Yang:2018nqn}.) For multiplicatively renormalizable nonlocal correlators such as the quasi-PDFs given above, the renormalization is similar but now one requires calculating the renormalization factors at each $z$.

The quark and gluon quasi-PDFs can, in general, mix with each other under renormalization.
In Ref.~\cite{Zhang:2018diq}, we have argued that inserting the gluon quasi-PDF operator into a quark state only yields finite mixing as long as all subdivergences have been renormalized (note the difference from the quark and gluon lightcone PDF operators which mix with each other under renormalization~\cite{Balitsky:1987bk,Braun:2019qtp}). The mixing effect can, in principle, be deferred to be considered at the factorization stage. Here we find that taking into account the mixing at the renormalization stage will help improve the convergence in the implementation of the matching in the RI/MOM scheme. To this end, it suffices to consider the following mixing of quasi-PDFs
\begin{eqnarray}\label{ZZ}
\begin{pmatrix}
 O^{(n)}_{g}(z,0) \\  O^s_{q}(z,0)
\end{pmatrix}
&=&
\begin{pmatrix}
Z_{11}(z) & Z_{12}(z)/z \\ z Z_{21}(z) & Z_{22}(z)
\end{pmatrix}
\begin{pmatrix}
 O^{(n)}_{g,R}(z,0) \\   O^s_{q,R}(z,0)
\end{pmatrix},
\end{eqnarray}
where $O^s_q(z_1, z_2)=1/2[\bar q_i (z_1) \Gamma W(z_1,z_2)  q_i(z_2)-(z_1\leftrightarrow z_2)]$ is the $C$-even combination of quark operators, $Z_{ij}(z)$ are dimensionless factors, and $z$ compensates for the different mass dimension between the quark and gluon quasi-PDF operators. In the limit $z\to 0$ (taken after combining the entries of the mixing matrix and the operators), the above mixing pattern reduces to the mixing pattern of local operators.

The renormalization factors in the above mixing matrix can be determined using the following renormalization conditions
\begin{align}
\label{renormcond}
\frac{{\rm Tr} [\Lambda_{22}(p,z){\cal P}]_R}{{\rm Tr} [\Lambda_{22}(p,z){\cal P}]_{\rm tree}}\bigg|_{\tiny\begin{matrix}p^2=-\mu_R^2 \\ \!\!\!\!p_z=p^R_z\end{matrix}}&=1,  \hspace{2em}
\frac{[P_{ij}^{ab}\Lambda_{11}^{ab,ij}(p,z)]_R}{[P_{ij}^{ab} \Lambda_{11}^{ab,ij}(p,z)]_{\rm tree}}\bigg|_{\tiny\begin{matrix}p^2=-\mu_R^2 \\ \!\!\!\!p_z=p^R_z\end{matrix}}=1,\nonumber\\
{\rm Tr} [\Lambda_{12}(p,z){\cal P}]_R\bigg|_{\tiny\begin{matrix}p^2=-\mu_R^2 \\ \!\!\!\!p_z=p^R_z\end{matrix}}&=0, \hspace{2em}
{[P_{ij}^{ab}\Lambda_{21}^{ab,ij}(p,z)]_R}\bigg|_{\tiny\begin{matrix}p^2=-\mu_R^2 \\ \!\!\!\!p_z=p^R_z\end{matrix}}=0,
\end{align}
where $\Lambda_{\{11,12\}}$ ($\Lambda_{\{21,22\}}$) denote the amputated Green's functions of $O_{g}^{(n)}$ ($O^s_{q}$) in an offshell gluon and quark state, respectively.
$\cal P$ and $P_{ij}^{ab}$ are projection operators that are associated with the quark and gluon matrix elements and define the RI/MOM renormalization factors. $\mu_R$ and $p_z^R$ are unphysical scales introduced in the RI/MOM scheme to specify the subtraction point. $b, c$ are color indices and $i,j$ Lorentz indices. In the non-singlet quark PDF case with $\Gamma=\gamma^t$~\cite{Liu:2018uuj}, the amputated Green's function has the following structure
\begin{equation}\label{quarkME}
\Lambda_{\gamma^t}(p, z)= \widetilde{f}_t(p,z)\gamma^t  + \widetilde{f}_{z}(p,z)\frac{p^t\gamma^z}{p^z} +\widetilde{f}_p(p,z)\frac{p^t\slashed{p}}{p^2},
\end{equation}
and $\cal P$ was chosen there in such a way that it projects out the coefficient of $\gamma^t$ only, which captures all terms in $\Lambda_{\gamma^t}(p,z)$ that lead to UV divergences in the local limit. However, in general both the coefficient of $\gamma^t$ and $\gamma^z$ can lead to UV divergences in the local limit. This is the case, e.g., in the mixing diagram to be considered below. We will need both coefficients to define the RI/MOM counterterm. As for $P_{ij}^{ab}$, a simple choice is $P_{ij}^{ab}= \delta^{ab}g_{\perp,ij}/(2-D)$, where $g_{\perp,ij}$ denotes the transverse metric tensor and $D$ is the spacetime dimension.

Defining the inverse of the renormalization matrix in Eq.~(\ref{ZZ}) as
\begin{align}\label{ZZinverse}
{\cal {\bar Z}}&=\begin{pmatrix}
{\bar Z}_{11}(z) & {\bar Z}_{12}(z)/z \\ z {\bar Z}_{21}(z) & {\bar Z}_{22}(z)
\end{pmatrix}=\begin{pmatrix}
Z_{11}(z) & Z_{12}(z)/z \\ z Z_{21}(z) & Z_{22}(z)
\end{pmatrix}^{-1},
\end{align}
we then have from Eqs.~(\ref{ZZ}), (\ref{renormcond}) and (\ref{ZZinverse})
\begin{align}
{\bar Z}_{11}(z)=\frac{{[P_{ij}^{ab} \Lambda_{11}^{ab,ij}(p,z)]_{\rm tree}}{\rm Tr} [\Lambda_{22}(p,z){\cal P}]}{([P_{ij}^{ab} \Lambda_{11}^{ab,ij}(p,z)]{\rm Tr} [\Lambda_{22}(p,z){\cal P}]-[P_{ij}^{ab}\Lambda_{21}^{ab,ij}(p,z)]{\rm Tr} [\Lambda_{12}(p,z){\cal P}])}\bigg|_{\tiny\begin{matrix}p^2=-\mu_R^2 \\ \!\!\!\!p_z=p^R_z\end{matrix}},\nonumber\\
{\bar Z}_{12}(z)/z=-\frac{{[P_{ij}^{ab} \Lambda_{11}^{ab,ij}(p,z)]_{\rm tree}}{\rm Tr} [\Lambda_{12}(p,z){\cal P}]}{([P_{ij}^{ab} \Lambda_{11}^{ab,ij}(p,z)]{\rm Tr} [\Lambda_{22}(p,z){\cal P}]-[P_{ij}^{ab}\Lambda_{21}^{ab,ij}(p,z)]{\rm Tr} [\Lambda_{12}(p,z){\cal P}])}\bigg|_{\tiny\begin{matrix}p^2=-\mu_R^2 \\ \!\!\!\!p_z=p^R_z\end{matrix}},\nonumber\\
z {\bar Z}_{21}(z)=-\frac{{[P_{ij}^{ab} \Lambda_{21}^{ab,ij}(p,z)]}{\rm Tr} [\Lambda_{22}(p,z){\cal P}]_{\rm tree}}{([P_{ij}^{ab} \Lambda_{11}^{ab,ij}(p,z)]{\rm Tr} [\Lambda_{22}(p,z){\cal P}]-[P_{ij}^{ab}\Lambda_{21}^{ab,ij}(p,z)]{\rm Tr} [\Lambda_{12}(p,z){\cal P}])}\bigg|_{\tiny\begin{matrix}p^2=-\mu_R^2 \\ \!\!\!\!p_z=p^R_z\end{matrix}},\nonumber\\
{\bar Z}_{22}(z)=\frac{{[P_{ij}^{ab} \Lambda_{11}^{ab,ij}(p,z)]}{\rm Tr} [\Lambda_{22}(p,z){\cal P}]_{\rm tree}}{([P_{ij}^{ab} \Lambda_{11}^{ab,ij}(p,z)]{\rm Tr} [\Lambda_{22}(p,z){\cal P}]-[P_{ij}^{ab}\Lambda_{21}^{ab,ij}(p,z)]{\rm Tr} [\Lambda_{12}(p,z){\cal P}])}\bigg|_{\tiny\begin{matrix}p^2=-\mu_R^2 \\ \!\!\!\!p_z=p^R_z\end{matrix}}.
\end{align}

Denoting the hadron matrix element of $O(z, 0)$ as $h(z, P^z, 1/a)$, i.e., $h_i(z, P^z, 1/a)=\langle P|O_i(z,0)|P\rangle$, $i=q, g$, the renormalized hadron matrix elements then read
\begin{eqnarray}
h_{g,R}^{(n)}(z,P^z,\mu_R, p_z^R)&=& {\bar Z}_{11}(z,\mu_R, p_z^R,1/a) h_{g}^{(n)}(z,P^z,1/a)+ {\bar Z}_{12}(z,\mu_R, p_z^R,1/a)/z~ h_{q}^{s}(z,P^z,1/a), \nonumber \\
h_{q,R}^{s}(z,P^z,\mu_R, p_z^R)&=& {\bar Z}_{22}(z,\mu_R, p_z^R,1/a)h_{q}^{s}(z,P^z,1/a) + z {\bar Z}_{21}(z,\mu_R, p_z^R,1/a)~ h_{g}^{(n)}(z,P^z,1/a).
\end{eqnarray}
The renormalized quasi-PDF in the RI/MOM scheme can be obtained from the above renormalized matrix elements by a Fourier transform given in Eqs.~(\ref{qlcDef}) and (\ref{gpdfdef}), respectively.
Note that we can take the continuum limit $a\to0$ in ${h}_R$ since all terms singular in $a$ have been removed by the renormalization procedure. This means that the factorization of the renormalized matrix element can be studied in the continuum, as will be done in the next subsection.

\subsection{Factorization}
In Ref.~\cite{Zhang:2018diq}, we have given a general factorization formula for the quark and gluon quasi-PDFs in the presence of mixing. In this subsection, we give a detailed derivation of it using the operator product expansion (OPE), along the same line as that used for the isovector quark quasi-PDF~\cite{Izubuchi:2018srq}. For illustration purposes, we choose $\Gamma=\gamma^t$ for the quark quasi-PDF and $O^{(4)}_{g}$ for the gluon quasi-PDF. The derivation for other operators follows straightforwardly from what is presented below.

The renormalized quark and gluon nonlocal operator matrix elements can be expanded in terms of gauge-invariant local operator matrix elements to the leading-twist approximation as
\begin{align}
{\tilde h}_{q_i,R}(z, P^z, \mu)&\simeq\frac{1}{2P^t}\sum_{n=1}^\infty \frac{(-iz)^{n-1}}{(n-1)!}\Big[C^{(n-1)}_{q_i q_j}(\mu^2 z^2)\langle P|n^t_{\mu_1} n_{\mu_2}...n_{\mu_n}O_{q_j}^{\mu_1...\mu_n}(\mu)|P\rangle\nonumber\\
&+C^{(n-1)}_{q g}(\mu^2 z^2)\langle P|n^t_{\mu_1} n_{\mu_2}...n_{\mu_n} O_g^{\mu_1...\mu_n}(\mu)|P\rangle\Big],\nonumber\\
{\tilde h}_{g,R}(z, P^z, \mu)&\simeq\frac{1}{(P^z)^2}\sum_{n=2}^\infty \frac{(-iz)^{n-2}}{(n-2)!}\Big[C^{(n-2)}_{gg}(\mu^2 z^2)\langle P|n_{\mu_1}...n_{\mu_n}O_{g}^{\mu_1...\mu_n}(\mu)|P\rangle\nonumber\\
&+C^{(n-2)}_{g q}(\mu^2 z^2)\langle P|n_{\mu_1}...n_{\mu_n} O_{q_j}^{\mu_1...\mu_n}(\mu)|P\rangle\Big],
\end{align}
where we have introduced extra normalization factors so that the two matrix elements have the same mass dimension. For simplicity, we have also denoted all renormalization scales with $\mu$. $n^{t,\rho}=(1,0,0,0)$ and $n^\rho=(0,0,0,1)$, $C^{(n)}_{q_i q_j}=\delta_{ij}+\frac{\alpha_s}{2\pi}C_{q_i q_j}^{(n),1}+{\cal O}(\alpha_s^2)$, $C^{(n)}_{\{q g, g q\}}=\frac{\alpha_s}{2\pi}C_{\{qg, gq\}}^{(n),1}+{\cal O}(\alpha_s^2)$ and $C_{gg}^{(n)}=1+\frac{\alpha_s}{2\pi}C_{gg}^{(n), 1}+{\cal O}(\alpha_s^2)$ denote the Wilson coefficients. $O_{q_j}^{\mu_1...\mu_n}$ and $O_g^{\mu_1...\mu_n}$ are the renormalized symmetric traceless twist-2 quark and gluon operators
\begin{align}
O_{q_j}^{\mu_1...\mu_n}&= Z^{n}_{q_j}\big[{\bar q}_j(0) \gamma^{\{\mu_1}iD^{\mu_2}\cdots iD^{\mu_n\}}q_j(0)-{\rm trace}\big],\nonumber\\
O_g^{\mu_1...\mu_n}&=Z_g^{n}\big[F^{\{\mu_1\nu}(0)iD^{\mu_2}\cdots iD^{\mu_{n-1}}F_\nu^{\,\mu_n\}}(0)-{\rm trace}\big],
\end{align}
where $\{\cdots\}$ denotes a symmetrization of the enclosed indices. Their matrix elements are related to the moments of quark and gluon PDF, respectively
\begin{align}
\langle P|O_{q_j}^{\mu_1...\mu_n}|P\rangle&=2a_{q_j, n}(\mu)(P^{\mu_1}\cdots P^{\mu_n}-{\rm trace}),\nonumber\\
\langle P|O_g^{\mu_1...\mu_n}|P\rangle&=2a_{g, n}(\mu)(P^{\mu_1}\cdots P^{\mu_n}-{\rm trace}),
\end{align}
with
\begin{eqnarray}
a_{q_j, n}(\mu)=\int_{-1}^1 dx \, x^{n-1} f_{q_j/H}(x,\mu), \hspace{2em} a_{g, n}(\mu)=\frac 1 2 \int_{-1}^1 dx \, x^{n-1} f_{g/H}(x,\mu).
\end{eqnarray}
Owing to the symmetry of the gluon PDF, $a_{g,n}$ does not vanish only for even $n$.

Let us first consider ${\tilde h}_{q_i,R}(z, P^z, \mu)$. Ignoring all trace terms, we can write
\begin{align}
{\tilde h}_{q_i,R}(z, P^z, \mu)&=\sum_{n=1}^\infty \frac{(-iz P^z)^{n-1}}{(n-1)!}\Big[C^{(n-1)}_{q_i q_j}(\mu^2 z^2)a_{q_j, n}(\mu)+C^{(n-1)}_{q g}(\mu^2 z^2)a_{g, n}(\mu)\Big]\nonumber\\
&=\sum_{n=1}^\infty \frac{(-i\nu)^{n-1}}{(n-1)!}\Big[C^{(n-1)}_{q_i q_j}(\mu^2 z^2)\int_{-1}^1 dx \, x^{n-1} f_{q_j/H}(x,\mu)+\frac{C^{(n-1)}_{q g}(\mu^2 z^2)}{2}\int_{-1}^1 dx \, x^{n-1} f_{g/H}(x,\mu)\Big]\nonumber\\
&=\sum_{n=1}^\infty \frac{(-i\nu)^{n-1}}{(n-1)!}C^{(n-1)}_{q_i q_j}(\mu^2 z^2)\int_{-1}^1 dx \, x^{n-1} \int \frac{d\nu'}{2\pi} e^{ix\nu'}h_{q_j}(\nu',\mu)\nonumber\\
&+\sum_{n=2, {\rm even}}^\infty \frac{(-i\nu)^{n-1}}{(n-1)!}\frac{C^{(n-1)}_{q g}(\mu^2 z^2)}{2}\int_{-1}^1 dx \, x^{n-2} \int \frac{d\nu'}{2\pi} e^{ix\nu'}h_{g}(\nu',\mu),
\end{align}
where we have introduced the Ioffe-time $\nu=-z\cdot P=z P^z$ and $\nu'=-\xi\cdot P=-P^+\xi^-$, $h_{q_i/g, R}$ denote the coordinate space matrix elements used to define the quark and gluon PDFs at lightlike separation $\xi^2=0$. Defining
\begin{align}
\int \frac{d\nu}{2\pi}e^{iu\nu}\sum_{n=1}^\infty \frac{(-i\nu)^{n-1}}{(n-1)!}C^{(n-1)}_{q_i q_j}(\mu^2 z^2)&={\cal C}_{q_i q_j}(u, \mu^2 z^2),\nonumber\\
\int \frac{d\nu}{2\pi}e^{iu\nu}\sum_{n=2, {\rm even}}^\infty \frac{(-i\nu)^{n-2}}{(n-1)!}\frac{C^{(n-1)}_{qg}(\mu^2 z^2)}{2}&=i\,{\cal C}_{qg}(u, \mu^2 z^2),
\end{align}
with $u$ being in the range $(-1, 1)$~\cite{Radyushkin:2016hsy,Radyushkin:2017cyf}, we then have
\begin{align}
{\tilde h}_{q_i,R}(z, P^z, \mu)&=\int_{-1}^1 dx\int_{-1}^1 du\, e^{-i u x \nu}\bigg[{\cal C}_{q_i q_j}(u, \mu^2 z^2) \int \frac{d\nu'}{2\pi} e^{ix\nu'}h_{q_j}(\nu',\mu)
+\nu{\cal C}_{qg}(u, \mu^2 z^2) \int \frac{d\nu'}{2\pi} e^{ix\nu'}h_{g}(\nu',\mu)\bigg]\nonumber\\
&=\int_{-1}^1 du\, {\cal C}_{q_i q_j}(u, \mu^2 z^2)h_{q_j}(u \nu,\mu)+\int_{-1}^1 du\, \nu{\cal C}_{qg}(u, \mu^2 z^2) h_{g}(u \nu,\mu).
\end{align}
This is the general factorization of the coordinate space matrix element in the presence of mixing. To convert it to the factorization of quasi-PDFs, we need a Fourier transform of the above relation
\begin{align}
&{\tilde f}_{q_i/H}(x, P^z,\mu)=P^z\int\frac{dz}{2\pi}e^{i z x P^z}{\tilde h}_{q_i,R}(z, P^z, \mu)\nonumber\\
&=P^z\int\frac{dz}{2\pi}e^{i z x P^z}\sum_{n=1}^\infty \frac{(-iz P^z)^{n-1}}{(n-1)!}\Big[C^{(n-1)}_{q_i q_j}(\mu^2 z^2)\int_{-1}^1 dy \, y^{n-1} f_{q_j/H}(y)+\frac{C^{(n-1)}_{q g}(\mu^2 z^2)}{2}\int_{-1}^1 dy \, y^{n-1} f_{g/H}(y)\Big]\nonumber\\
&=\int_{-1}^1 dy\int dz' \delta(z'-z y)\int\frac{dz P^z}{2\pi}e^{i z x P^z}\sum_{n=1}^\infty \frac{(-iz y P^z)^{n-1}}{(n-1)!}\Big[C^{(n-1)}_{q_i q_j}(\mu^2 z^2) f_{q_j/H}(y)+\frac{C^{(n-1)}_{q g}(\mu^2 z^2)}{2} f_{g/H}(y)\Big]\nonumber\\
&=\int_{-1}^1 \frac{dy}{|y|}\int \frac{d\nu'}{2\pi}\,e^{i\nu' x/y}\sum_{n=1}^\infty \frac{(-i\nu')^{n-1}}{(n-1)!}\Big[C^{(n-1)}_{q_i q_j}\Big(\frac{\mu^2 {\nu'}^2}{y^2(P^z)^2}\Big) f_{q_j/H}(y)+\frac{1}{2}C^{(n-1)}_{q g}\Big(\frac{\mu^2 {\nu'}^2}{y^2(P^z)^2}\Big) f_{g/H}(y)\Big]\nonumber\\
&=\int_{-1}^1 \frac{dy}{|y|}\Big[C_{q_i q_j}\Big(\frac{x}{y}, \frac{\mu}{y P^z}\Big) f_{q_j/H}(y)+C_{qg}\Big(\frac{x}{y}, \frac{\mu}{y P^z}\Big) f_{g/H}(y)\Big],
\end{align}
where we have defined
\begin{align}
C_{q_i q_j}\Big(\frac{x}{y}, \frac{\mu}{y P^z}\Big)&=\int \frac{d\nu'}{2\pi}\,e^{i\nu' x/y}\sum_{n=1}^\infty \frac{(-i\nu')^{n-1}}{(n-1)!}C^{(n-1)}_{q_i q_j}\Big(\frac{\mu^2 {\nu'}^2}{y^2(P^z)^2}\Big),\nonumber\\
C_{qg}\Big(\frac{x}{y}, \frac{\mu}{y P^z}\Big)&=\int \frac{d\nu'}{2\pi}\,e^{i\nu' x/y}\sum_{n=1}^\infty \frac{(-i\nu')^{n-1}}{(n-1)!}C^{(n-1)}_{qg}\Big(\frac{\mu^2 {\nu'}^2}{y^2(P^z)^2}\Big)/2.
\end{align}

Now let us turn to ${\tilde h}_{g,R}(z, P^z, \mu)$. By ignoring all trace terms, one can write as before
\begin{align}
{\tilde h}_{g,R}(z, P^z, \mu)&=\sum_{n=2}^\infty \frac{(-i\nu)^{n-2}}{(n-2)!}\Big[C^{(n-2)}_{gg}(\mu^2 z^2)\int_{-1}^1 dx \, x^{n-1} f_{g/H}(x,\mu)+2C^{(n-2)}_{gq}(\mu^2 z^2)\int_{-1}^1 dx \, x^{n-1} f_{q_i/H}(x,\mu)\Big]\nonumber\\
&=\sum_{n=2, \rm{even}}^\infty \frac{(-i\nu)^{n-2}}{(n-2)!}C^{(n-2)}_{gg}(\mu^2 z^2)\int_{-1}^1 dx \, x^{n-2} \int \frac{d\nu'}{2\pi} e^{ix\nu'}h_{g}(\nu',\mu)\nonumber\\
&+\sum_{n=2}^\infty \frac{(-i\nu)^{n-2}}{(n-2)!}2C^{(n-2)}_{gq}(\mu^2 z^2)\int_{-1}^1 dx \, x^{n-1} \int \frac{d\nu'}{2\pi} e^{ix\nu'}h_{q_i}(\nu',\mu).
\end{align}
Defining
\begin{align}
\int \frac{d\nu}{2\pi}e^{iu\nu}\sum_{n=2, \rm{even}}^\infty \frac{(-i\nu)^{n-2}}{(n-2)!}C^{(n-2)}_{gg}(\mu^2 z^2)&={\cal C}_{gg}(u, \mu^2 z^2),\nonumber\\
\int \frac{d\nu}{2\pi}e^{iu\nu}\sum_{n=2}^\infty \frac{(-i\nu)^{n-1}}{(n-2)!}2C^{(n-2)}_{gq}(\mu^2 z^2)&=-i\,{\cal C}_{gq}(u, \mu^2 z^2),
\end{align}
we then have the following factorization in coordinate space
\begin{align}
{\tilde h}_{g,R}(z, P^z, \mu)&=\int_{-1}^1 dx\int_{-1}^1 du\, {e^{-i u x \nu}}\bigg[{\cal C}_{gg}(u, \mu^2 z^2) \int \frac{d\nu'}{2\pi} e^{ix\nu'}h_{g}(\nu',\mu)
+\frac{{\cal C}_{gq}(u, \mu^2 z^2)}{\nu} \int \frac{d\nu'}{2\pi} e^{ix\nu'}h_{q_i}(\nu',\mu)\bigg]\nonumber\\
&=\int_{-1}^1 du\, {\cal C}_{gg}(u, \mu^2 z^2) h_{g}(u \nu,\mu)+\int_{-1}^1 du \frac{{\cal C}_{gq}(u, \mu^2 z^2)}{\nu} h_{q_i}(u \nu,\mu).
\end{align}
The factorization in momentum space reads
\begin{align}
&{\tilde f}_{g/H}(x, P^z,\mu)=P^z\int\frac{dz}{2\pi x}e^{i z x P^z}{\tilde h}_{g,R}(z, P^z, \mu)\nonumber\\
&=\int\frac{dz P^z}{2\pi x}e^{i z x P^z}\sum_{n=2}^\infty \frac{(-iz P^z)^{n-2}}{(n-2)!}\Big[C^{(n-2)}_{gg}(\mu^2 z^2)\int_{-1}^1 dy \, y^{n-1} f_{g/H}(y)+2C^{(n-2)}_{gq}(\mu^2 z^2)\int_{-1}^1 dy \, y^{n-1} f_{q_i/H}(y)\Big]\nonumber\\
&=\int_{-1}^1 \frac{dy}{|y|}\frac{y}{x}\int \frac{d\nu'}{2\pi}\,e^{i\nu' x/y}\sum_{n=2}^\infty \frac{(-i\nu')^{n-2}}{(n-2)!}\Big[C^{(n-2)}_{gg}\Big(\frac{\mu^2 {\nu'}^2}{y^2(P^z)^2}\Big) f_{g/H}(y)+2C^{(n-2)}_{gq}\Big(\frac{\mu^2 {\nu'}^2}{y^2(P^z)^2}\Big) f_{q_i/H}(y)\Big]\nonumber\\
&=\int_{-1}^1 \frac{dy}{|y|}\Big[C_{gq}\Big(\frac{x}{y}, \frac{\mu}{y P^z}\Big) f_{q_j/H}(y)+C_{gg}\Big(\frac{x}{y}, \frac{\mu}{y P^z}\Big) f_{g/H}(y)\Big],
\end{align}
where we have defined
\begin{align}
C_{gg}\Big(\frac{x}{y}, \frac{\mu}{y P^z}\Big)&=\frac{y}{x}\int \frac{d\nu'}{2\pi}\,e^{i\nu' x/y}\sum_{n=2}^\infty \frac{(-i\nu')^{n-2}}{(n-2)!}C^{(n-2)}_{gg}\Big(\frac{\mu^2 {\nu'}^2}{y^2(P^z)^2}\Big),\nonumber\\
C_{gq}\Big(\frac{x}{y}, \frac{\mu}{y P^z}\Big)&=\frac{y}{x}\int \frac{d\nu'}{2\pi}\,e^{i\nu' x/y}\sum_{n=2}^\infty \frac{(-i\nu')^{n-2}}{(n-2)!}2C^{(n-2)}_{gq}\Big(\frac{\mu^2 {\nu'}^2}{y^2(P^z)^2}\Big).
\end{align}

Restoring all renormalization scales, the general factorization of the quark and gluon quasi-PDFs reads
\begin{align}\label{allfac}
\tilde f^{(n)}_{g/H}(x ,P^z, p_z^R, \mu_R)=&\int_{-1}^1\frac{dy}{|y|}\Big[C_{gg}\Big(\frac{x}{y}, \frac{\mu_R}{p_z^R},\frac{y P^z}{\mu}, \frac{y P^z}{p_z^R}\Big)f_{g/H}(y, \mu)+C_{gq}\Big(\frac{x}{y}, \frac{\mu_R}{p_z^R},\frac{y P^z}{\mu}, \frac{y P^z}{p_z^R}\Big)f_{q_j/H}(y, \mu)\Big]\nonumber\\
&+\mathcal O\Big(\frac{M^2}{(P^z)^2}, \frac{\Lambda_{\rm QCD}^2}{(P^z)^2}\Big),\nonumber\\
\tilde f_{q_i/H}(x ,P^z, p_z^R, \mu_R)=&\int_{-1}^1\frac{dy}{|y|}\Big[C_{q_i q_j}\Big(\frac{x}{y}, \frac{\mu_R}{p_z^R},\frac{y P^z}{\mu}, \frac{y P^z}{p_z^R}\Big)f_{q_j/H}(y, \mu)+C_{qg}\Big(\frac{x}{y}, \frac{\mu_R}{p_z^R},\frac{y P^z}{\mu}, \frac{y P^z}{p_z^R}\Big)f_{g/H}(y, \mu)\Big]\nonumber\\
&+\mathcal O\Big(\frac{M^2}{(P^z)^2}, \frac{\Lambda_{\rm QCD}^2}{(P^z)^2}\Big),
\end{align}
where a summation of $j$ over all quark flavors is implied. The factorization for the polarized quasi-PDFs has the same form as Eq.~(\ref{allfac}) with all unpolarized distributions being replaced by the polarized ones and also different hard coefficients. It is worthwhile to point out that the higher-twist contributions shall behave like $1/[x^2(1-x) (P^z)^2]$ instead of $1/(P^z)^2$, as demonstrated in Ref.~\cite{Braun:2018brg}.

\section{One-loop matching for unpolarized quasi-PDFs in RI/MOM scheme}
\label{sec:1loopmatching_unpolarized}

As shown in the previous section, when the hadron momentum $P^z$ is much larger than the hadronic scale, the higher-twist contributions get suppressed (except for very small/large $x$), the quasi-PDFs can be factorized into the lightcone PDFs with perturbatively calculable hard matching coefficients. In this section, we present the one-loop calculation of the hard matching coefficients for unpolarized quark and gluon quasi-PDFs in the presence of mixing. The polarized case will be discussed in the next section. Our result is obtained in the RI/MOM scheme, which can be used to connect the RI/MOM renormalized quasi-PDFs to the PDFs in $\overline{\rm MS}$ scheme. Since the matching depends on UV physics only and not on the external state, we can calculate it in quark or gluon external states $|q(p)\rangle$, $|g(p)\rangle$. The infrared (IR) divergences can be regularized using their offshellness.

\subsection{Gluon in Gluon}

Let us start with the gluon matrix element of the gluon quasi-PDF operator, which is the most complicated among all calculations. At tree-level one finds:
\begin{eqnarray}
x \tilde f^{(n,0)}_{g/g}(x, \rho) =\delta(x-1), \;\;\;
x   f^{(0)}_{g/g}(x, \mu) =\delta(x-1),
\end{eqnarray}
where $\rho=(-p^2-i\epsilon)/p_z^2$ and $i\epsilon$ allows for an analytic continuation from $\rho<1$ to $\rho>1$. As before, we have ignored the crossed terms which can be obtained from $\{\tilde f, f\}(x)=-\{\tilde f, f\}(-x)$. Ignoring such terms has no impact on the extraction of the matching coefficient. The above results lead to the following tree-level matching coefficient:
\begin{eqnarray}
C^{(0)}_{gg}(x/y)= \delta(x/y-1).
\end{eqnarray}

At one-loop level, the partonic quasi-PDF can be written as follows:
\begin{eqnarray}\label{1loopsep}
x \tilde f_{g/g}^{(n)}(x,\rho)=  \big[x \tilde f_{g/g}^{(n)}(x,\rho)\big]_+ +  \tilde c^{(n)}\delta(x-1),
\end{eqnarray}
with $n=1,2,3,4$, and the ``+" subscript denotes the usual plus-prescription
\begin{eqnarray}
[f(x)]_+=f(x)-\delta(1-x)\int dx'\, f(x').
\end{eqnarray}
Integrating Eq.~(\ref{1loopsep}) over the momentum fraction, one arrives at
\begin{eqnarray}
 \int dx\, x  \tilde f_{g/g}^{(n)}(x)=  \tilde c^{(n)},
\end{eqnarray}
which corresponds to the matrix element of local operators
\begin{eqnarray}
 \tilde   c^{(n)} &=& \frac{1}{p_z^2} N^{(n)} \langle g(p)|O_g^{(n)}(0, 0) |g(p)\rangle.
\end{eqnarray}

Before we proceed, a few general remarks on the calculation below are in order.
\begin{itemize}
\item The above equations apply to   bare operator matrix elements. One can write down similar equations for the renormalized ones. In our calculation of the matching coefficients, the PDF is renormalized in ${\rm \overline {MS}}$ scheme while the quasi-PDF is renormalized in the RI/MOM scheme. The renormalized local operator matrix elements in these two schemes differ from each other in general.

\item The offshell gluon matrix elements of gauge invariant operators can mix with those of gauge variant operators.
To illustrate this point, it is worthwhile to consider the UV divergence from the offshell gluon matrix element of the local gluon operator $F^{\mu\alpha}(0)F^{\nu\beta}(0)$:
\begin{eqnarray}
&& \langle p, \rho|F^{\mu\alpha}(0)F^{\nu\beta}(0) |p, \sigma\rangle = -   \frac{\alpha_s C_A}{4\pi\epsilon} \nonumber\\
&& \times \Bigg\{
\frac{1}{12} p^2 \bigg(9 {g}^{\alpha \nu } {g}^{\beta \sigma } {g}^{\mu \rho }-9 {g}^{\alpha \beta } {g}^{\mu \rho } {g}^{\nu \sigma }-{g}^{\alpha \nu } {g}^{\beta \rho } {g}^{\mu \sigma }+{g}^{\alpha \beta } {g}^{\mu \sigma } {g}^{\nu \rho }+{g}^{\alpha \sigma } \left({g}^{\beta \rho } {g}^{\mu \nu }-{g}^{\beta \mu } {g}^{\nu \rho }\right)\nonumber\\
&&+{g}^{\alpha \rho } \left(9 {g}^{\beta \mu } {g}^{\nu \sigma }-9 {g}^{\beta \sigma } {g}^{\mu \nu }\right)-2 {g}^{\alpha \nu } {g}^{\beta \mu } {g}^{\rho \sigma }+2 {g}^{\alpha \beta } {g}^{\mu \nu } {g}^{\rho \sigma }\bigg)\nonumber\\
&& +
\frac{1}{6} {p}^{\mu } {p}^{\nu } \left(4 {g}^{\alpha \sigma } {g}^{\beta \rho }+10 {g}^{\alpha \rho } {g}^{\beta \sigma }-7 {g}^{\alpha \beta } {g}^{\rho \sigma }\right)-\frac{1}{6} {p}^{\beta } {p}^{\mu } \left(4 {g}^{\alpha \sigma } {g}^{\nu \rho }+10 {g}^{\alpha \rho } {g}^{\nu \sigma }-7 {g}^{\alpha \nu } {g}^{\rho \sigma }\right)\nonumber\\
&&-\frac{1}{6} {p}^{\alpha } {p}^{\nu } \left(10 {g}^{\beta \sigma } {g}^{\mu \rho }+4 {g}^{\beta \rho } {g}^{\mu \sigma }-7 {g}^{\beta \mu } {g}^{\rho \sigma }\right)+\frac{1}{6} {p}^{\alpha } {p}^{\beta } \left(4 {g}^{\mu \sigma } {g}^{\nu \rho }+10 {g}^{\mu \rho } {g}^{\nu \sigma }-7 {g}^{\mu \nu } {g}^{\rho \sigma }\right)\nonumber\\
&&
-\frac{3}{4} {p}^{\mu } {p}^{\rho } \left({g}^{\alpha \nu } {g}^{\beta \sigma }-{g}^{\alpha \beta } {g}^{\nu \sigma }\right)-\frac{3}{4} {p}^{\nu } {p}^{\sigma } \left({g}^{\alpha \rho } {g}^{\beta \mu }-{g}^{\alpha \beta } {g}^{\mu \rho }\right)+\frac{3}{4} {p}^{\alpha } {p}^{\rho } \left({g}^{\beta \sigma } {g}^{\mu \nu }-{g}^{\beta \mu } {g}^{\nu \sigma }\right)\nonumber\\
&&+\frac{3}{4} {p}^{\beta } {p}^{\sigma } \left({g}^{\alpha \rho } {g}^{\mu \nu }-{g}^{\alpha \nu } {g}^{\mu \rho }\right)-\frac{1}{6} {p}^{\mu } {p}^{\sigma } \left({g}^{\alpha \nu } {g}^{\beta \rho }-{g}^{\alpha \beta } {g}^{\nu \rho }\right)-\frac{1}{6} {p}^{\nu } {p}^{\rho } \left({g}^{\alpha \sigma } {g}^{\beta \mu }-{g}^{\alpha \beta } {g}^{\mu \sigma }\right)\nonumber\\
&&+\frac{1}{6} {p}^{\alpha } {p}^{\sigma } \left({g}^{\beta \rho } {g}^{\mu \nu }-{g}^{\beta \mu } {g}^{\nu \rho }\right)+\frac{1}{6} {p}^{\beta } {p}^{\rho } \left({g}^{\alpha \sigma } {g}^{\mu \nu }-{g}^{\alpha \nu } {g}^{\mu \sigma }\right)+\frac{1}{6} {p}^{\rho } {p}^{\sigma } \left({g}^{\alpha \nu } {g}^{\beta \mu }-{g}^{\alpha \beta } {g}^{\mu \nu }\right)\Bigg\}+{\cal O}(\epsilon^0),\label{eq:general_divergence}
\end{eqnarray}
with the crossed contributions being neglected.
This leads to the following contributions to the UV divergences in $ \tilde c^{(n)}$:
\begin{eqnarray}\label{localren}
 \tilde   c^{(1,g)}&=& \frac{\alpha_s C_A}{12\pi \epsilon }\frac{p^2}{p^2+p_z^2}+{\cal O}({\epsilon^0}),\nonumber\\
 \tilde   c^{(2,g)} &=& -\frac{\alpha_s C_A}{12\pi \epsilon }\frac{p^2}{p_z^2}+{\cal O}({\epsilon^0}),\nonumber\\
  \tilde   c^{(3,g)} &=&{\cal O}({\epsilon^0}),\nonumber\\
  \tilde   c^{(4,g)} &=&\frac{\alpha_s C_A}{3\pi \epsilon }\frac{p^2}{p_z^2}+{\cal O}({\epsilon^0}),
\end{eqnarray}
if a physical projection ${P}^{ab}_{ij}= \delta^{ab} g_{\perp,ij}/(2-D)$ is employed. As can be seen from the above equations, the UV divergences might depend on the offshellness of external gluons, which is a sign of the potential mixing with gauge variant operators. It is interesting to note that the UV divergence of $\tilde c^{(3,g)}$ is independent of $p^2$. This is because it corresponds to the $tz$ component of the gluon energy momentum tensor for which all gauge-variant operators to mix turn out to vanish~\cite{Collins:1994ee,Yang:2018bft}. As we will see below, such a behavior is consistent with the asymptotic behavior at large $x$ of the quasi-PDF defined with $O_{g,R}^{(3)}(z,0)$, which does not depend on $p^2$ either. This feature turns out to help achieve a better convergence in the implementation of the matching. Thus, in the following we will focus on $O_{g,R}^{(3)}(z,0)$,  and present the one-loop matching calculation for the gluon quasi-PDF defined with this operator. For completeness and comparison purposes, the results for other definitions are also collected in Appendix \ref{sec:oneloopRxi}.

\item In pure Yang-Mills theory, $O_g^{(3)}(0,0)$ does not renormalize, as shown by the results in Eq.~(\ref{localren})~\footnote{In general, one should be cautious about offshell gluons, as calculating the matrix element of gluon energy momentum tensor in offshell gluon states and then taking the onshell limit is rather tricky due to the existence of IR divergences~\cite{Collins:1994ee,Harris:1994tp}.}. In QCD, quarks can enter the gluon diagrams relevant for the above calculation, but only through gluon wave function renormalization at one-loop level, and lead to the following contribution to ${\tilde c}^{(3,g)}$ and ${c}^{(3,g)}$ (the counterpart of ${\tilde c}^{(3,g)}$ for the gluon PDF) after renormalization
\begin{eqnarray}\label{gglocalterm}
{\tilde c}^{(3, g)}_{\rm RI/MOM}=   1-  \frac{\alpha_s T_f}{3\pi} \left(  -\ln\frac{-p^2}{\mu^2_R}  \right), \;\;\;
  c^{3, g}_{{\rm \overline {MS}}}=   1-  \frac{\alpha_s T_f}{3\pi} \left(  -\ln\frac{-p^2}{\mu^2}+ \frac{5}{3} \right) .
\end{eqnarray}
This will be needed in the calculation of the matching coefficient below.

\end{itemize}

\begin{figure}[tbp]
\centering
\includegraphics[width=0.5\textwidth]{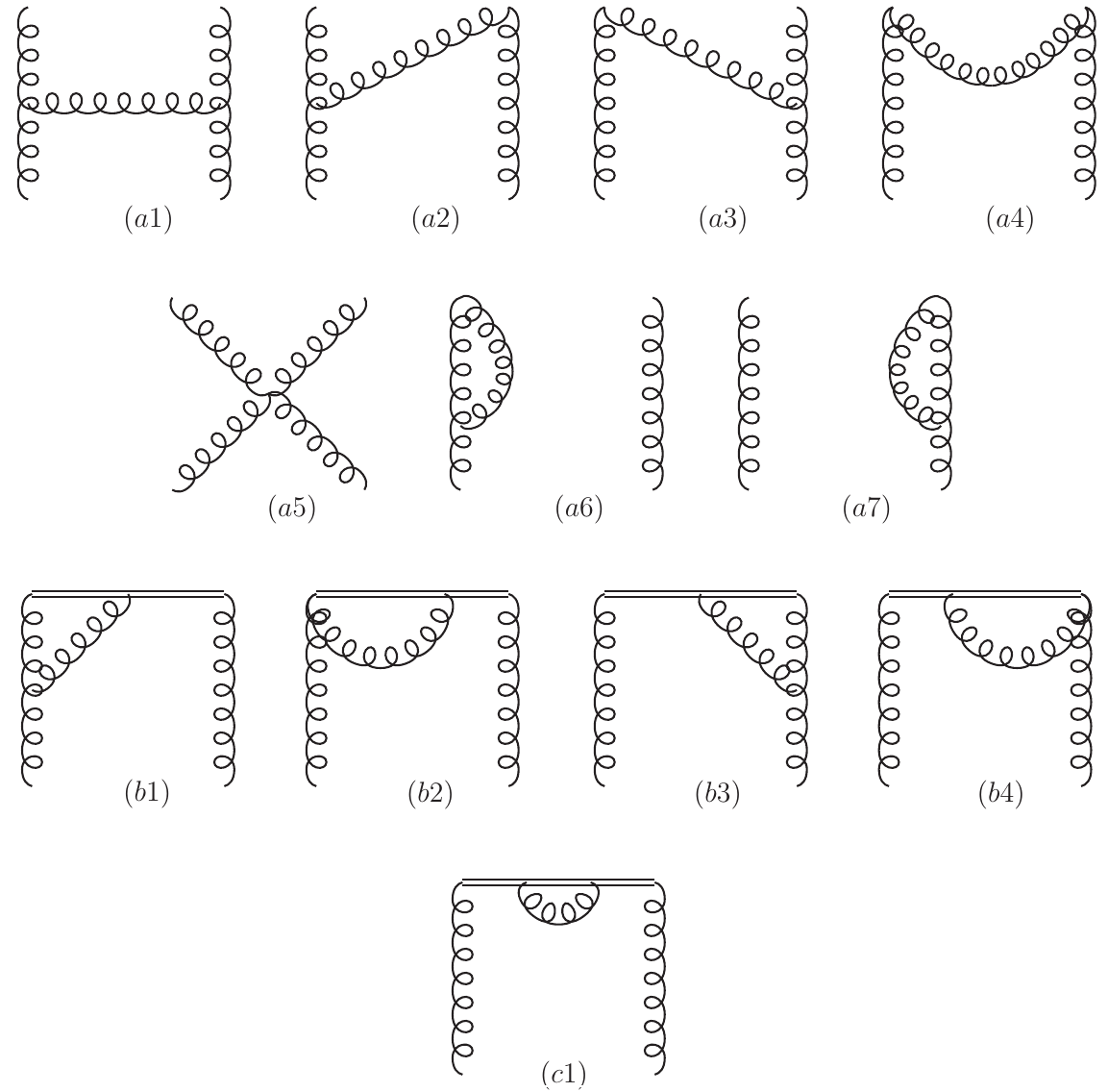}
\caption{One-loop diagrams for the gluon quasi-PDF. The gluon self-energy diagrams are not shown. } \label{fig:gluon2gluon}
\end{figure}

Now we present the one-loop results for the partonic quasi-PDF and PDF. The calculation is carried out in Landau gauge, and the steps are similar to those presented in Refs.~\cite{Wang:2017qyg,Wang:2017eel}. Given Eqs.~(\ref{1loopsep}) and (\ref{gglocalterm}), we only present the distribution part, {\it i.e.} the first term in Eq.~(\ref{1loopsep}). To this end, we need to calculate the one-loop matrix element of $O_g^{(3)}(z,0)$ in an offshell gluon state. The relevant Feynman diagrams are shown in Fig.~\ref{fig:gluon2gluon}, and the result reads
 \begin{align}
     &\big[x\tilde f_{g/g}^{(3,1)}(x, \rho)\big]_+\nonumber\\
   =&\frac{\alpha_s C_A}{2\pi}  \left\{ \begin{array}{ll} \bigg[\frac{-(\rho -4)^2 (\rho -1)+8 (\rho +2) x^4-16 (\rho +2) x^3-2 \left(\rho ^2+8 \rho -24\right) x^2+\left(6 \rho ^2+20 \rho -32\right) x}{8 (\rho -1)^2 (x-1)}\frac{1}{\sqrt{1-\rho}}\ln \frac{2x-1-\sqrt{1-\rho}}{2x-1+\sqrt{1-\rho}} \\
  +\frac{4 x^3}{(2 x-1) \left(\rho +4 x^2-4 x\right)}+\frac{8 x^4-16 x^3-22 x^2+34 x-9}{4 (\rho -1) (x-1) (2 x-1)}-\frac{8 x^3(x-1)}{\left(\rho +4 x^2-4 x\right)^2}+\frac{3 (2
   x-1) x}{2 (\rho -1)^2}-\frac{4 x+1}{4 (x-1)}\bigg]_+,\;\;&x>1\\
  \bigg[\frac{-(\rho -4)^2 (\rho -1)+8 (\rho +2) x^4-16 (\rho +2) x^3-2 \left(\rho ^2+8 \rho -24\right) x^2+\left(6 \rho ^2+20 \rho -32\right) x}{8 (\rho -1)^2 (x-1)}\frac{1}{\sqrt{1-\rho}}\ln \frac{1-\sqrt{1-\rho}}{1+\sqrt{1-\rho}}\\
  +\frac{-30 x^2+34 x-9}{4 (\rho -1) (x-1)}+\frac{3 \left(4 x^3-4 x^2+x\right)}{2 (\rho -1)^2}+\frac{6 x+1}{4 (x-1)}\bigg]_+,\;\;&0<x<1\\
  \bigg[-\frac{-(\rho -4)^2 (\rho -1)+8 (\rho +2) x^4-16 (\rho +2) x^3-2 \left(\rho ^2+8 \rho -24\right) x^2+\left(6 \rho ^2+20 \rho -32\right) x}{8 (\rho -1)^2 (x-1)}\frac{1}{\sqrt{1-\rho}}\ln \frac{2x-1-\sqrt{1-\rho}}{2x-1+\sqrt{1-\rho}}\\
  -\frac{4 x^3}{(2 x-1) \left(\rho +4 x^2-4 x\right)}+\frac{-8 x^4+16 x^3+22 x^2-34 x+9}{4 (\rho -1) (x-1) (2 x-1)}+\frac{8 x^3(x-1)}{\left(\rho +4 x^2-4 x\right)^2}-\frac{3
   (2 x-1) x}{2 (\rho -1)^2}+\frac{4 x+1}{4 (x-1)}\bigg]_+,\;\;&x<0.\end{array}\right.
  \end{align}
{It is worthwhile to point out that the leading logarithmic terms in the $\rho\to 0$ limit are consistent with those presented in Ref.~\cite{Wang:2017qyg}. Similar agreement also exists in the results presented below.}
As in the quark case~\cite{Stewart:2017tvs,Liu:2018uuj}, the bare quasi-PDF result is obtained by taking the onshell limit $\rho\to0$ of the above expression except where it has to be kept as an IR regulator
  \begin{eqnarray}
 \big[ x\tilde f_{g/g}^{(3,1)}(x, \rho\to 0)\big]_+ &=\frac{\alpha_s C_A}{2\pi} \left\{ \begin{array}{ll} \bigg[\frac{2(1-x+x^2)^2}{x-1}\ln\frac{x-1}{x}+\frac{4x^3-6x^2+8x-5}{2(x-1)}\bigg]_+,\;\;&x>1\\
  \bigg[\frac{2(1-x+x^2)^2}{x-1}\ln\frac{\rho}{4}+\frac{12x^4-24x^3+30x^2-17x+5}{2(x-1)}\bigg]_+,\;\;&0<x<1\\
 \bigg[-\frac{2(1-x+x^2)^2}{x-1}\ln\frac{x-1}{x}-\frac{4x^3-6x^2+8x-5}{2(x-1)}\bigg]_+,\;\;&x<0.\end{array}\right.
  \end{eqnarray}

The renormalized lightcone PDF can be calculated analogously, and gives
\begin{align}
\bigg[   x f_{g/g}^{(1)}\bigg(x, \frac{\mu^2}{-p^2}\bigg)\bigg]_+&=\theta(x)\theta(1-x)\bigg\{\frac{\alpha_s C_A}{2\pi}\bigg[\frac{2(1-x+x^2)^2}{x-1}\ln\frac{-p^2 x(1-x)}{\mu^2}+2x^3-2x^2+3x-2 \bigg]_+  \nonumber\\
& \;\;\; - \frac{\alpha_s C_A}{4\pi} \bigg[\frac{x}{1-x}\bigg]_+\bigg\},
\end{align}
where the result in the first square bracket is the same as the Feynman gauge result.

The one-loop matching coefficient is given by the difference in the renormalized quasi-PDF and lightcone PDF
\begin{eqnarray}
x C^{(3,1)}_{gg}(x,r, \frac{p_z}{\mu}, \frac{p_z}{p_z^R})&=& \bigg[x\tilde f_{g/g}^{(3,1)}(x, \rho\to 0)   - x  f_{g/g}^{(1)}\bigg(x, \frac{\mu^2}{-p^2}\bigg)- (x\tilde f_{g/g}^{(3,1)})_{C.T.}\bigg]_+ + \Big({\tilde c}^{(3,g)}_{{\rm RI/MOM}}-c^{3, g}_{\overline {\rm MS}}\Big)\delta(x-1),
\end{eqnarray}
where the $\ln (-p^2)$ dependence in each individual term cancels out in the combination on the r.h.s., and the counterterm in the RI/MOM scheme can be determined from the renormalization condition above as
\begin{eqnarray}
(x\tilde f_{g/g}^{(3,1)})_{C.T.}= \left|\frac{p_z}{p_z^R}\right|  x\tilde f_{g/g}^{(3,1)}\left(\frac{p_z}{p_z^R}(x-1)+1, r\right)
\end{eqnarray}
with $r=\mu_R^2/(p_z^R)^2$.

\subsection{Quark in Quark}

\begin{figure}[tbp]
\centering
\includegraphics[width=0.5\textwidth]{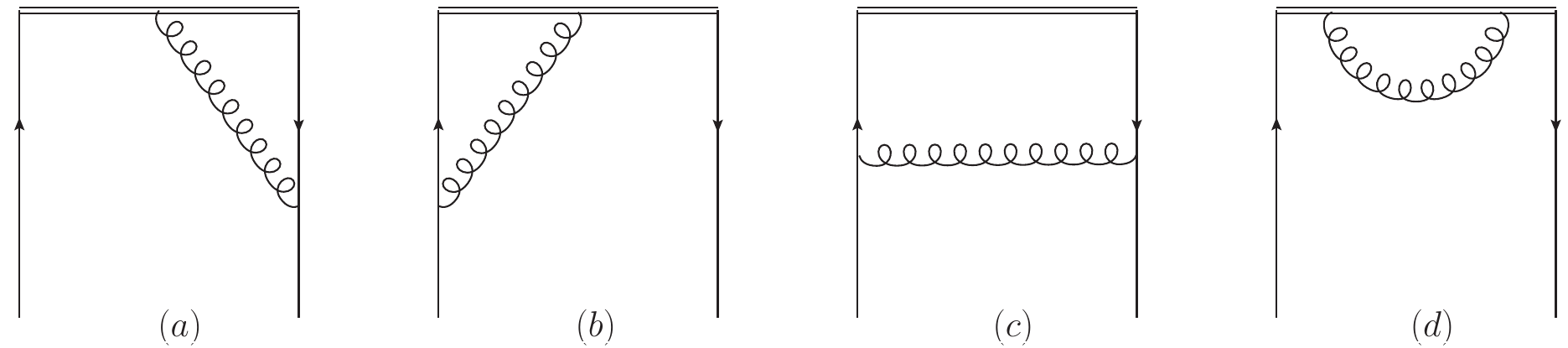}
\caption{One-loop diagrams for the quark quasi-PDF. The quark self-energy diagrams are not shown. } \label{fig:quark2quark}
\end{figure}

This case has already been considered at one-loop in Ref.~\cite{Liu:2018uuj}. For completeness, we also quote the results here and briefly explain how it was obtained. As we will see below, our definition of the counterterm differs from that defined in Ref.~\cite{Liu:2018uuj} by a finite piece. The relevant Feynman diagrams are shown in Fig.~\ref{fig:quark2quark}.

Owing to the offshellness of the external quark, the one-loop quark quasi-PDF contains two more Dirac structures apart from the tree-level one $\gamma^t$, and is given by the following projection~\cite{Liu:2018uuj}
\begin{eqnarray}
{\rm Tr}\left[ \left(  \left[\tilde f_{q/q,t}^{(1)}(x,\rho) \right]_+\gamma^t +  \left[\tilde f_{q/q,z}^{(1)}(x,\rho) \right]_+ \frac{p^t}{p^z} \gamma^z + \left[\tilde f_{q/q,p}^{(1)}(x,\rho) \right]_+ \frac{p^t p\!\!\!\slash}{p^2}\right) {\cal P}\right],
\end{eqnarray}
where the coefficients of $\gamma^t$ and $\gamma^z$ read in Landau gauge
\begin{eqnarray}
  \tilde f_{q/q,t}^{(1)}(x, \rho)
  &=&\frac{\alpha_s C_F}{2\pi} \left\{ \begin{array}{ll} \frac{2 x^2}{(2 x-1) \left(\rho +4 x^2-4 x\right)}+\frac{4 x-3}{2 (\rho -1) (2 x-1)}-\frac{3}{2 (x-1)}-\frac{\left(3 \rho +4 x^2+(\rho -8) x\right) \ln \frac{2 x-1-\sqrt{1-\rho }}{2 x-1+\sqrt{1-\rho }}}{4 (1-\rho )^{3/2} (x-1)},\;\;&x>1\\
  \frac{4 x-3}{2 (\rho -1)}+\frac{3}{2 (x-1)}-\frac{\ln\frac{1-\sqrt{1-\rho }}{1+\sqrt{1-\rho}} \left(3 \rho +4 x^2+(\rho -8) x\right)}{4 (1-\rho )^{3/2} (x-1)},\;\;&0<x<1\\
  -\frac{2 x^2}{(2 x-1) \left(\rho +4 x^2-4 x\right)}+\frac{3-4 x}{2 (\rho -1) (2 x-1)}+\frac{3}{2 (x-1)}+\frac{\left(3 \rho +4 x^2+(\rho -8) x\right) \ln \frac{2 x-1-\sqrt{1-\rho }}{2 x-1+\sqrt{1-\rho }}}{4 (1-\rho )^{3/2} (x-1)},\;\;&x<0, \\
  \end{array}\right.\\
  \tilde f_{q/q,z}^{(1)}(x, \rho)  &=&\frac{\alpha_s C_F}{2\pi}  \left\{ \begin{array}{ll} \frac{\left(2 \rho ^2+3 \rho +4 (\rho +2) x^2-(13 \rho +8) x+4\right) \ln \frac{2 x-1-\sqrt{1-\rho }}{2 x-1+\sqrt{1-\rho }}}{4 (1-\rho )^{5/2} (x-1)}
  +\frac{2 \left(3 x^2-2 x\right)}{(2 x-1)^3 \left(\rho +4 x^2-4 x\right)}\\-\frac{8 \left(x^3-x^2\right)}{(2 x-1) \left(\rho +4 x^2-4 x\right)^2}+\frac{8 x^4-34 x^3+40 x^2-17 x+2}{(\rho -1) (x-1) (2 x-1)^3}+\frac{3 (4 x-3)}{2 (\rho -1)^2 (2 x-1)},\;\;&x>1\\
  \frac{\ln \frac{1-\sqrt{1-\rho }}{1+\sqrt{1-\rho }}  \left(2 \rho ^2+3 \rho +4 (\rho +2) x^2-(13 \rho +8) x+4\right)}{4 (1-\rho )^{5/2} (x-1)}
  +\frac{2-3 x}{(\rho -1) (x-1)}+\frac{3 (4 x-3)}{2 (\rho -1)^2},\;\;&0<x<1\\
  -\frac{\left(2 \rho ^2+3 \rho +4 (\rho +2) x^2-(13 \rho +8) x+4\right) \ln \frac{2 x-1-\sqrt{1-\rho }}{2 x-1+\sqrt{1-\rho }}}{4 (1-\rho )^{5/2} (x-1)}
  -\frac{2 \left(3 x^2-2 x\right)}{(2 x-1)^3 \left(\rho +4 x^2-4 x\right)}\\+\frac{8 \left(x^3-x^2\right)}{(2 x-1) \left(\rho +4 x^2-4 x\right)^2}+\frac{-8 x^4+34 x^3-40 x^2+17 x-2}{(\rho -1) (x-1) (2 x-1)^3}-\frac{3 (4 x-3)}{2 (\rho -1)^2 (2 x-1)},\;\;&x<0.
  \end{array}\right.\end{eqnarray}

In Ref.~\cite{Liu:2018uuj}, a so-called minimal projector for $\cal P$ has been used, which determines the bare quark quasi-PDF as
\begin{eqnarray}
  \Big[\tilde f_{q/q}^{(1)}(x, \rho\to 0)\Big]_+ = \left[\tilde f_{q/q,t}^{(1)}(x, \rho\to 0) \right]_++  \left[\tilde f_{q/q,z}^{(1)}(x, \rho\to 0) \right]_+,
\end{eqnarray}
with the following explicit form
\begin{eqnarray}
  \Big[\tilde f_{q/q}^{(1)}(x, \rho\to 0)\Big]_+&=\frac{\alpha_s C_F}{2\pi} \left\{ \begin{array}{ll}\left[ \frac{x^2+1}{x-1} \ln\frac{x-1}{x}+1\right]_+,\;\;&x>1\\
 \left[ \frac{x^2+1}{x-1} \ln \frac{\rho}{4}+\frac{8 x^2-8 x+5}{2 (x-1)}\right]_+,\;\;&0<x<1\\
 -\left[\frac{x^2+1}{x-1} \ln\frac{x-1}{x}+1\right]_+,\;\;&x<0.\end{array}\right.\end{eqnarray}
Note that there is no extra local term like ${\tilde c}^{(3,g)}_{\rm RI/MOM}$ above due to vector current conservation. The renormalized lightcone quark PDF has the following expression
\begin{eqnarray}
   \Big[f_{q/q}^{(1)}\Big(x, \frac{\mu^2}{-p^2}\Big)\Big]_+  =
 \bigg\{\frac{\alpha_s C_F}{2\pi}   \left[\frac{x^2+1}{x-1}  \ln  \frac{-p^2  (1-x) x}{\mu^2}+\frac{-5+10x-6x^2}{2(1-x) }\right]_+\bigg\}\theta(x)\theta(1-x),
\end{eqnarray}

The matching coefficient can then be extracted as
\begin{eqnarray}
C^{(1)}_{qq}\bigg(x,r, \frac{p_z}{\mu}, \frac{p_z}{p_z^R}\bigg)&=& \bigg[\tilde f_{q/q}^{(1)}(x, \rho\to 0)   -    f_{q/q}^{(1)}\bigg(x, \frac{\mu^2}{-p^2}\bigg)- (\tilde f_{q/q}^{(1)})_{C.T.}\bigg]_+,
\end{eqnarray}
where again the $\ln (-p^2)$ dependence cancels out in the combination on the r.h.s., and the counterterm in the RI/MOM scheme is given by
\begin{eqnarray}
(\tilde f_{q/q}^{(1)})_{C.T.}= \left|\frac{p_z}{p_z^R}\right|  \tilde f_{q/q, t}^{(1)}\left(\frac{p_z}{p_z^R}(x-1)+1, r\right)+\left|\frac{p_z}{p_z^R}\right|  \tilde f_{q/q, z}^{(1)}\left(\frac{p_z}{p_z^R}(x-1)+1, r\right).
\end{eqnarray}
Note that the counterterm defined here differs from that given in Ref.~\cite{Liu:2018uuj} by a finite piece. In Ref.~\cite{Liu:2018uuj} the projector used to define the counterterm was chosen differently from that used to define the bare quasi-PDF, and projected out the coefficient of $\gamma^t$ only since only this coefficient contributes to $1/|x|$ in the asymptotic limit $x\to\infty$. In the present paper, we use the same projector to determine the counterterms in the quark matrix elements of quark and gluon quasi-PDF operators. As can be seen from the next subsection, for the latter we need $\cal P$ to project out both the coefficients of $\gamma^t$ and $\gamma^z$. Therefore the same projection shall apply to the former. In fact, projecting out both coefficients is more natural, since in the infinite momentum limit both $\gamma^t$ and $\gamma^z$ approach $\gamma^+$, therefore both may contribute to UV divergences. From a different point of view, we can always rewrite $\gamma^z$ in terms of $\gamma^t$ and $\slashed p$ if the external quark has no transverse momentum. This also implies that taking both the coefficients of $\gamma^t$ and $\gamma^z$ to define the counterterm is more natural.

\subsection{Gluon in Quark}

\begin{figure}[tbp]
\centering
\includegraphics[width=0.1\textwidth]{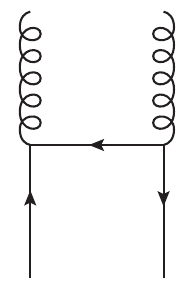}
\caption{One-loop diagram for the quark matrix element of the gluon quasi-PDF operator.   } \label{fig:quark2gluon}
\vspace*{-.5em}
\end{figure}

Now we turn to the mixing contributions. Let us first consider the quark matrix element of the gluon quasi-PDF operator, whose one-loop diagram is given in Fig.~\ref{fig:quark2gluon}.

To illustrate the kinematic dependence of the mixing terms, it is useful to begin with the one-loop quark matrix element of the local operator $F^{\mu\alpha}(0) F^{\nu\beta}(0)$
\begin{eqnarray}
 \langle p|F^{\mu\alpha}(0) F^{\nu\beta}(0)|p\rangle &=&  -\frac{\alpha_s C_F}{12\pi \epsilon} \bar u(p)  \bigg(-{\gamma}^{\mu } {p}^{\beta } {g}^{\alpha \nu }+{\gamma}^{\alpha } {p}^{\beta } {g}^{\mu \nu }+{\gamma}^{\beta } {p}^{\alpha } {g}^{\mu \nu }-{\gamma}^{\beta } {p}^{\mu } {g}^{\alpha \nu }\nonumber\\
&&
+{\gamma}^{\nu } \left({p}^{\mu } {g}^{\alpha \beta }-{p}^{\alpha } {g}^{\beta \mu }\right)+{\gamma}^{\mu } {p}^{\nu } {g}^{\alpha \beta }-{\gamma}^{\alpha } {p}^{\nu } {g}^{\beta \mu }+ p\!\!\!\slash\left({g}^{\alpha \nu } {g}^{\beta \mu }-{g}^{\alpha \beta } {g}^{\mu \nu }\right)\bigg)u(p).
\end{eqnarray}
From the above result we obtain
\begin{eqnarray}\label{mixkin}
&& \langle p|O_{g,R}^{(3)}(0, 0)|p\rangle =  \frac{\alpha_s C_F}{6\pi \epsilon} \bar u(p) [p^t \gamma^z + p^z \gamma^t] u(p) +{\cal O}(\epsilon^0).
\label{eq:quark2gluon_local_unpolarized_divergence}
\end{eqnarray}
As the $tz$ component of the gluon energy momentum tensor, $O_{g,R}^{(3)}(0,0)$ in general mixes with the same component of the quark contribution
\begin{eqnarray}
T^{tz}_q= \frac{1}{2} i \bar \psi i D^{(t} \gamma^{z)} \psi +  \frac{1}{2} i \bar \psi i {\overset{\leftarrow}{D}}\!\;^{(t} \gamma^{z)} \psi,
\end{eqnarray}
where $(\cdots)$ denotes an antisymmetrization of the enclosed indices. The above operator has the same momentum dependence as Eq.~(\ref{mixkin}) when sandwiched in a quark state. This indicates that the mixing matrix element in Fig.~\ref{fig:quark2gluon} has the same momentum dependence as the tree-level quark contribution, which is indeed needed to define an appropriate RI/MOM counterterm.

The renormalized mixing contribution from the lightcone gluon PDF has the following form
\begin{align}
  xf_{g/q}^{(1)}\bigg(x, \frac{\mu^2}{-p^2}\bigg)=\frac{\alpha_s C_F}{2\pi}\bigg[(1+(1-x)^2)\ln\frac{\mu^2}{-p^2 x(1-x)}+x(1-x)-2\bigg].
\end{align}

For the quasi-PDF, we follow the decomposition as in the quark case:
\begin{eqnarray}
{\rm Tr}\left[ \left( x f_{g/q,t}^{(3,1)}(x,\rho) \gamma^t + xf_{g/q,z}^{(3,1)}(x,\rho) \frac{p^t}{p^z} \gamma^z +x  f_{g/q,p}^{(3,1)}(x,\rho) \frac{p^t p\!\!\!\slash}{p^2}\right) {\cal P}\right],
\end{eqnarray}
and choose the projector ${\cal P}$ such that it projects out the coefficients of both $\gamma^t$ and $\gamma^z$. We therefore have
\begin{eqnarray}
x \tilde f_{g/q}^{(3,1)}  = x f_{g/q,t}^{(3,1)}+x  f_{g/q,z}^{(3,1)},
\end{eqnarray}
leading to
 \begin{align}
  x\tilde f_{g/q}^{(3,1)}(x, \mu, P^z)  &=\frac{\alpha_s C_F}{2\pi}  \left\{ \begin{array}{ll} -\frac{5 \rho ^2-10 \rho +(8 \rho +4) x^2-4 (\rho +2) x+8}{4 (\rho -1)^{2}}\frac{1}{\sqrt{1-\rho}}\ln \frac{2x-1-\sqrt{1-\rho}}{2x-1+\sqrt{1-\rho}} \\
   -\frac{(\rho -4) \rho +8 (2 \rho +1) x^3-4 \left(\rho ^2+2 \rho +6\right) x^2+2 \left(3 \rho ^2-2 \rho +8\right) x}{2 (1-\rho )^{2} \left(\rho +4 x^2-4 x\right)},\;\;&x>1\\
  -\frac{5 \rho ^2-10 \rho +(8 \rho +4) x^2-4 (\rho +2) x+8}{4 (\rho -1)^{2}}\frac{1}{\sqrt{1-\rho}}\ln \frac{1-\sqrt{1-\rho}}{1+\sqrt{1-\rho}}\\
  -\frac{(2 x-1) (\rho +2 (\rho +2) x-4)}{2 (1-\rho )^{2}},\;\;&0<x<1\\
  \frac{5 \rho ^2-10 \rho +(8 \rho +4) x^2-4 (\rho +2) x+8}{4 (\rho -1)^{2}}\frac{1}{\sqrt{1-\rho}}\ln \frac{2x-1-\sqrt{1-\rho}}{2x-1+\sqrt{1-\rho}} \\
  +\frac{(\rho -4) \rho +8 (2 \rho +1) x^3-4 \left(\rho ^2+2 \rho +6\right) x^2+2 \left(3 \rho ^2-2 \rho +8\right) x}{2 (1-\rho )^{2} \left(\rho +4 x^2-4 x\right)},\;\;&x<0.
  \end{array}  \right.
  \end{align}
  In the limit $\rho\to 0$, we have for the bare quasi-PDF
  \begin{align}
  x\tilde f_{g/q}^{(3,1)}(x, \mu, P^z)  &=\frac{\alpha_s C_F}{2\pi} \left\{ \begin{array}{ll}
  - \left(1+(1-x)^2\right) \ln \frac{x-1}{x}- x+2,\;\;&x>1\\
  -\left(1+(1-x)^2\right) \ln\frac{\rho}{4}-4 x^2+6 x-2,\;\;&0<x<1\\
 \left(1+(1-x)^2\right) \ln \frac{x-1}{x}+x-2,\;\;&x<0.\\
  \end{array}\right.,
  \end{align}
In the limit $x\to \infty$, the above expression behaves asymptotically as
\begin{eqnarray}
  x\tilde f_{g/q}^{(3,1)}(x, \mu, P^z) \to \frac{\alpha_sC_F}{2\pi} \left(\frac{1}{2} + \frac{4}{3x}\right).
\end{eqnarray}
If one integrates over the momentum fraction with DR, it is straightforward to see that the above behavior is consistent with the local result in Eq.~(\ref{eq:quark2gluon_local_unpolarized_divergence}).

As before, the matching coefficient can be extracted as
\begin{eqnarray}
x C_{g/q}^{(3,1)}\bigg(x,r, \frac{p_z}{\mu}, \frac{p_z}{p_z^R}\bigg)&=& \bigg[x \tilde f_{g/q}^{(3,1)}(x, \rho\to 0)   - x  f_{g/q}^{(1)}\bigg(x, \frac{\mu^2}{-p^2}\bigg)- (x\tilde f_{g/q}^{(3,1)})_{C.T.}\bigg],
\end{eqnarray}
where the counterterm in the RI/MOM scheme is determined as
\begin{eqnarray}
(x\tilde f_{g/q}^{(3,1)})_{C.T.}= \left|\frac{p_z}{p_z^R}\right|  x\tilde f_{g/q}^{(3,1)}\left(\frac{p_z}{p_z^R}(x-1)+1, r\right).
\end{eqnarray}

\subsection{Quark in Gluon}
%

\begin{figure}[tbp]
\centering
\includegraphics[width=0.1\textwidth]{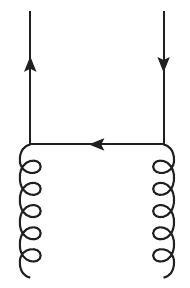}
\caption{One-loop diagram for the gluon matrix element of the quark quasi-PDF operator.   } \label{fig:gluon2quark}
\vspace*{-.5em}
\end{figure}


Now let us consider the gluon matrix element of the quark quasi-PDF operator, and the one-loop diagram is shown in Fig.~\ref{fig:gluon2quark}. We again start with the local matrix element
\begin{eqnarray}
\langle g(p)| \bar \psi \gamma^\mu \psi|g(p)\rangle = \epsilon_\sigma \epsilon^*_\rho
\frac{\alpha_s \left(-2  {p}^{\mu }  {g}^{\rho \sigma }+ {g}^{\mu \sigma }  {p}^{\rho }+ {g}^{\mu \rho }  {p}^{\sigma }\right)}{12 \pi  \epsilon }.
\end{eqnarray}
If $\mu=t$ and physical polarizations are used for the external gluons, one has the result:
\begin{eqnarray}
\langle g(p)| \bar \psi \gamma^t \psi|g(p)\rangle  = \frac{\alpha_s \sqrt{p^2+p_z^2}}{6\pi \epsilon},
\end{eqnarray}
which also has the same momentum dependence as the gluon matrix element of $O_{g,R}^{(3)}(0,0)$.

For the lightcone PDF, the result of the mixing diagram in Fig.~\ref{fig:gluon2quark} reads
  \begin{align}
   f_{q/g}^{(1)}\bigg(x, \frac{\mu^2}{-p^2}\bigg)=\frac{\alpha_s T_f}{2\pi}\bigg[(x^2+(1-x)^2)\ln\frac{\mu^2}{-p^2 x(1-x)}-1\bigg],
  \end{align}
while for the quasi-PDF one has:
  \begin{align}
  &\tilde f_{q/g}^{(1)}(x, \rho)\nonumber \\
  =&\frac{\alpha_s T_f}{2\pi}  \left\{ \begin{array}{ll} -\frac{\rho ^2-2 \rho +4 (\rho +2) x^2-4 (\rho +2) x+4}{4 (1-\rho)^{2}}\frac{1}{\sqrt{1-\rho}}\ln \frac{2x-1-\sqrt{1-\rho}}{2x-1+\sqrt{1-\rho}}
  -\frac{(2 x-1) \left(-(\rho -4) \rho +4 (\rho +2) x^2-4 (\rho +2) x\right)}{2 (1-\rho )^{2} \left(\rho +4 x^2-4 x\right)} ,\;\;&x>1\\
  -\frac{\rho ^2-2 \rho +4 (\rho +2) x^2-4 (\rho +2) x+4}{4 (1-\rho )^{2}}\frac{1}{\sqrt{1-\rho}}\ln \frac{1-\sqrt{1-\rho}}{1+\sqrt{1-\rho}}-\frac{-\rho +12 x^2-12 x+4}{2 (1-\rho )^{2}} ,\;\;&0<x<1\\
  \frac{\rho ^2-2 \rho +4 (\rho +2) x^2-4 (\rho +2) x+4}{4 (1-\rho )^{2}}\frac{1}{\sqrt{1-\rho}}\ln \frac{2x-1-\sqrt{1-\rho}}{2x-1+\sqrt{1-\rho}}
  -\frac{(2 x-1) \left((\rho -4) \rho -4 (\rho +2) x^2+4 (\rho +2) x\right)}{2 (1-\rho )^{2} \left(\rho +4 x^2-4 x\right)},\;\;&x<0.\\
  \end{array}\right.
  \end{align}
Taking $\rho\to 0$ gives the bare quasi-PDF result
  \begin{align}
\tilde f_{q/g}^{(1)}(x, \rho\to0)=\frac{\alpha_s T_f}{2\pi}  \left\{ \begin{array}{ll}
  -(x^2+(1-x)^2)\ln\frac{x-1}{x}-2x+1 ,\;\;&x>1\\
  -(x^2+(1-x)^2)\ln\frac{\rho}{4}-6x^2+6x-2,\;\;&0<x<1\\
  (x^2+(1-x)^2)\ln\frac{x-1}{x}+2x-1,\;\;&x<0.\\
  \end{array}\right.
  \end{align}

The matching coefficient is then given by
\begin{eqnarray}
C^{(1)}_{qg}\bigg(x,r, \frac{p_z}{\mu}, \frac{p_z}{p_z^R}\bigg)&=& \bigg[\tilde f_{q/g}^{(1)}(x, \rho\to 0)   -    f_{q/g}^{(1)}\bigg(x, \frac{\mu^2}{-p^2}\bigg)- (\tilde f_{q/g}^{(1)})_{C.T.}\bigg],
\end{eqnarray}
with
\begin{eqnarray}
(\tilde f_{q/g}^{(1)})_{C.T.}= \left|\frac{p_z}{p_z^R}\right|  \tilde f_{q/g}^{(1)}\left(\frac{p_z}{p_z^R}(x-1)+1, r\right).
\end{eqnarray}

\section{One-loop matching for polarized quasi-PDFs in RI/MOM scheme}
\label{sec:1loopmatching_polarized}

\subsection{Gluon in Gluon}

Now we turn to the polarized case. The calculation can be done in complete analogy with that presented in the previous section. As demonstrated in Ref.~\cite{Zhang:2018diq}, to study the polarized gluon PDF
\begin{eqnarray}
 \Delta   f_{g/H}(x,\mu) =  i\epsilon_{\perp ij} \int \frac{d\xi^-}{2\pi x P^+} e^{-i\xi^- xP^+}  \langle P|F^{+i}(\xi^- n_+)
  {\cal W}(\xi^- n_+,0; L_{n_+})F^{j+}(0) |P\rangle,\label{eq:gluon_light_cone_polarized_PDF}
\end{eqnarray}
we may use the following three operators to define the corresponding quasi-PDF
\begin{align}
\Delta {O}_g^{1}(z, 0) &=   i\epsilon_{\perp,ij}  F^{ti}(z_2) {\cal W}(z_2,z_1) F^{tj}(z_1), \\
 \Delta{  O}_g^{2}(z,0) &=   i\epsilon_{\perp,ij}  F^{zi}(z_2) {\cal W}(z_2,z_1) F^{zj}(z_1),\\
\Delta {O}_g^{3}(z,0) &=   i \epsilon_{\perp,ij} F^{ti}(z_2) {\cal W}(z_2,z_1) F^{zj}(z_1),
\end{align}
where  $ \epsilon_{\perp,ij} $ is the two-dimensional  antisymmetric tensor:
\begin{eqnarray}
\epsilon_{\perp,ij}= \epsilon_{\mu\nu ij} n_t^{\mu} n^\nu,
\end{eqnarray}
with the convention $\epsilon^{0123}=1$, and $n_t^\mu=(1,0,0,0)$.   The projection operator for the polarized gluon quasi-PDF is chosen as:
\begin{eqnarray}
{\cal P}_{\perp,ij}= \frac{i}{D-2}  \epsilon_{\mu\nu ij} n^{\mu}_t n^\nu.
\end{eqnarray}

As before,  we decompose the polarized quasi-PDF as
\begin{eqnarray}
x \Delta \tilde f_{g/g}^{(n)}(x)=  [x \Delta \tilde f]_+ +  \Delta \tilde  c^{(n)}\delta(x-1).
\end{eqnarray}
Integrating over the momentum fraction
\begin{eqnarray}
 \int_{-\infty}^{\infty} dx\,  x \Delta \tilde f_{g/g}^{(n)}(x)=  \Delta\tilde c^{(n)},
\end{eqnarray}
one obtains  the matrix element of the corresponding local operators:
\begin{eqnarray}
\Delta \tilde   c^{(n)} &=&  \frac{1}{(p^z)^2} \Delta N^{(n)} \langle g(p)|\Delta O_{g, R}^{(n)}(0, 0) |g(p)\rangle,
\end{eqnarray}
with
\begin{eqnarray}
\Delta N^{(1)} = \frac{(p^z)^2}{(p^t)^2}, \;\;\Delta N^{(2)}=1, \;\; \Delta N^{(3)} = \frac{p^z}{p^t}.
\end{eqnarray}

The local matrix elements have the following UV divergence structure
\begin{eqnarray}
\Delta \tilde   c^{(1)} &=& -\frac{\alpha_s C_A(p^2+6(p^z)^2)}{24\pi\epsilon (p^2+(p^z)^2)}, \\
\Delta \tilde   c^{(2)} &=& -\frac{\alpha_s C_A(5p^2+6(p^z)^2)}{24\pi\epsilon (p^z)^2}, \\
\Delta \tilde   c^{(3)} &=& -\frac{\alpha_s C_A }{4\pi\epsilon }, \label{eq:polarized_local_operator_UV_divergence}
\end{eqnarray}
where only the UV divergence of $\Delta \tilde c^{(3)}$ does not depend on the external momentum. For the same reason as the unpolarized case, we choose $\Delta O^{(3)}_{g}$ to define the polarized gluon quasi-PDF and present the corresponding one-loop matching kernel below.

The light-cone PDF yields the following real contribution
\begin{eqnarray}
  x \Delta f_{g/g}^{(1)}(x, \mu)
   =\frac{\alpha_sC_A}{2\pi} \left\{\frac{x}{x-1}\left[ \left(4 x^2-6 x+4\right) \ln \frac{-p^2   (1-x) x}{\mu^2}+8 x^2-11 x+7 + \frac{(1-\xi ) }{2}\right]\right\}_+\theta(x)\theta(1-x), \end{eqnarray}
whereas the quasi-PDF gives
\begin{eqnarray}
 x\Delta \tilde f_{g/g}^{(3,1)}(x, \rho) &=&\frac{\alpha_s C_A}{2\pi}  \left\{ \begin{array}{ll} -\frac{\rho  \left(\rho ^2-3 \rho +8\right)+8 (\rho -4) x^3+8 \left(\rho ^2-\rho +6\right) x^2-2 \left(9 \rho ^2-10 \rho +16\right) x }{8 (1-\rho )^{5/2} (x-1)}\ln \frac{2 x-1-\sqrt{1-\rho }}{2 x-1+\sqrt{1-\rho }}\\
 +\frac{4 x^3}{(2 x-1) \left(\rho +4 x^2-4 x\right)}+\frac{-8 x^3-8 x^2+14 x-3}{4 (\rho -1) (x-1) (2 x-1)}\\-\frac{8 \left(x^4-x^3\right)}{\left(\rho +4 x^2-4 x\right)^2}+\frac{3 (2 x-1)}{2 (\rho -1)^2}-\frac{4 x+1}{4 (x-1)},\;\;&x>1\\
 -\frac{ \rho  \left(\rho ^2-3 \rho +8\right)+8 (\rho -4) x^3+8 \left(\rho ^2-\rho +6\right) x^2-2 \left(9 \rho ^2-10 \rho +16\right) x}{8 (1-\rho )^{5/2} (x-1)}\ln \frac{1-\sqrt{1-\rho }}{1+\sqrt{1-\rho }}\\
 +\frac{3 \left(4 x^2-4 x+1\right)}{2 (\rho -1)^2}+\frac{-16 x^3+8 x^2+6 x-3}{4 (\rho -1) (x-1)}+\frac{6 x+1}{4 (x-1)},\;\;&0<x<1\\
 \frac{\rho  \left(\rho ^2-3 \rho +8\right)+8 (\rho -4) x^3+8 \left(\rho ^2-\rho +6\right) x^2-2 \left(9 \rho ^2-10 \rho +16\right) x}{8 (1-\rho )^{5/2} (x-1)}\ln \frac{2 x-1-\sqrt{1-\rho }}{2 x-1+\sqrt{1-\rho }}\\
 -\frac{4 x^3}{(2 x-1) \left(\rho +4 x^2-4 x\right)}+\frac{8 x^3+8 x^2-14 x+3}{4 (\rho -1) (x-1) (2 x-1)}\\+\frac{8 \left(x^4-x^3\right)}{\left(\rho +4 x^2-4 x\right)^2}-\frac{3 (2 x-1)}{2 (\rho -1)^2}+\frac{4 x+1}{4 (x-1)},\;\;&x<0.
\\\end{array}\right.\end{eqnarray}
In the $\rho\to0$ limit, the above result gets simplified
{\begin{eqnarray}
  x\Delta \tilde f_{g/g}^{(3,1)}(x, \rho)  &=\frac{\alpha_s C_A}{2\pi} \left\{ \begin{array}{ll} \frac{8 x^2+4 \left(2 x^2-3 x+2\right) x \ln \frac{x-1}{x}-8 x+1}{2 (x-1)},\;\;&x>1\\\frac{4 \left(2 x^2-3 x+2\right) x \ln \frac{\rho}{4}+20 x^3-28 x^2+15 x-1}{2 (x-1)},\;\;&0<x<1\\-\frac{8 x^2+4 \left(2 x^2-3 x+2\right) x \ln \frac{x-1}{x}-8 x+1}{2 (x-1)},\;\;&x<0. \\\end{array}\right.\end{eqnarray}

The virtual contribution is the same as the unpolarized case, whereas the real contribution differs in the asymptotic limit $x\to \infty$ as
\begin{eqnarray}
x\Delta \tilde f^{(3,1)}_{g/g}(x,\rho)- x  \tilde f^{(3,1)}_{g/g}(x,\rho) \to  \frac{\alpha_s C_A}{2\pi} \left(\frac{2}{3}  - \frac{1}{2x}\right).
\end{eqnarray}
Integrating over $x$ in DR, this gives the UV divergence in  Eq.~\eqref{eq:polarized_local_operator_UV_divergence} as expected.

The matching kernel can be written using the matching kernel for the unpolarized gluon quasi-PDF as
\begin{eqnarray}
x \Delta C^{(3,1)}_{gg}\bigg(x,r, \frac{p_z}{\mu}, \frac{p_z}{p_z^R}\bigg)&=& x C^{(3,1)}_{gg}\bigg(x,r, \frac{p_z}{\mu}, \frac{p_z}{p_z^R}\bigg)+  \bigg[ \Big(x\Delta \tilde f^{(3,1)}_{g/g}(x,\rho\to0)- x  \tilde f^{(3,1)}_{g/g}(x,\rho\to0)\Big)  \nonumber\\
&& \;\;\;  - \Big(x\Delta   f^{(3,1)}_{g/g}\Big(x, \frac{\mu^2}{-p^2}\Big)- x    f^{(3,1)}_{g/g}\Big(x, \frac{\mu^2}{-p^2}\Big)\Big)  - (x \Delta \tilde f_{g/g}^{(3,1)})_{C.T.}\bigg] ,
\end{eqnarray}
where again the $\ln (-p^2)$ dependence in each individual term cancels out in the combination on the r.h.s., and the counterterm in the RI/MOM scheme is determined as:
\begin{eqnarray}
(x\Delta \tilde f_{g/g}^{(3,1)})_{C.T.}= \left|\frac{p_z}{p_z^R}\right|  \left[x \Delta \tilde f_{g/g}^{(3,1)}\left(\frac{p_z}{p_z^R}(x-1)+1, r\right)-x   \tilde f_{g/g}^{(3,1)}\left(\frac{p_z}{p_z^R}(x-1)+1, r\right)\right].
\end{eqnarray}

\subsection{Quark in Quark}

For completeness, we also give the result for the polarized quark quasi-PDF and PDF defined as following
\begin{eqnarray}
\label{eq:quasi_quark_polarized}
\Delta { \tilde  f}_{q_i/H}(x,\mu, P^z) &=& \frac{P^z}{P^t}\int \frac{dz}{4\pi} e^{iz  xP^z}  \langle P|\overline{q}_i (z)
   \gamma^z   \gamma_5 W(z, 0) q_i(0) |P\rangle, \label{eq:polarized_quasi_quark_pdf}\\
\Delta   f_{q_i/H}(x,\mu) &=& \int \frac{d\xi^-}{4\pi} e^{-i\xi^- xP^+}  \langle P|\overline{q}_i (\xi^-)
   \gamma^+   \gamma_5  W(\xi^- ,0)q_i (0) |P\rangle.
\end{eqnarray}
The result for the polarized quark PDF is the same as that for the unpolarized one:
\begin{eqnarray}
\Delta f_{q/q}^{(1)}\bigg(x, \frac{\mu^2}{-p^2}\bigg)=f_{q/q}^{(1)}\bigg(x, \frac{\mu^2}{-p^2}\bigg).
\end{eqnarray}

For the quasi-PDF,
the one-loop result  can be decomposed as
\begin{eqnarray}
{\rm Tr}\left[ \left(  \left[\Delta \tilde f_{q/q,t}^{(1)} \right]_+\gamma^t \frac{p^t}{p^z} +  \left[\Delta \tilde f_{q/q,z}^{(1)} \right]_+  \gamma^z + \left[\Delta\tilde  f_{q/q,p}^{(1)} \right]_+ \frac{p^t p\!\!\!\slash}{p^2}\right)\gamma_5 {\cal P}\right],
\end{eqnarray}
where we define $\cal P$ to project out both the coefficients of $\gamma^t\gamma_5$ and $\gamma^z\gamma_5$. For the Dirac matrix in Eq.~\eqref{eq:quasi_quark_polarized}, $\Delta \tilde  f_{q/q,t}^{(1)}$ vanishes, and $\Delta \tilde f_{q/q,z}^{(1)}$ is given as:
\begin{eqnarray}
\Delta  \tilde f_{q/q,z}^{(1)}(x, \rho) &=&\frac{\alpha_s C_F}{2\pi}  \left\{ \begin{array}{ll} -\frac{3 \rho -2 x^2-2 }{2 (1-\rho )^{3/2} (x-1)}\ln \frac{2 x-1-\sqrt{1-\rho }}{2 x-1+\sqrt{1-\rho }}\\
+ \frac{4 x^2}{(2 x-1) \left(\rho +4 x^2-4 x\right)}+\frac{1-2 x^2}{(\rho -1) (x-1) (2 x-1)}-\frac{8 \left(x^3-x^2\right)}{\left(\rho +4 x^2-4 x\right)^2}-\frac{3}{2 (x-1)},\;\;&x>1\\
-\frac{3 \rho -2 x^2-2}{2 (1-\rho )^{3/2} (x-1)}\ln \frac{1-\sqrt{1-\rho }}{1+\sqrt{1-\rho }}
+\frac{1-2 x^2}{(\rho -1) (x-1)}+\frac{3}{2 (x-1)},\;\;&0<x<1\\
-\frac{-3 \rho +2 x^2+2}{2 (1-\rho )^{3/2} (x-1)}\ln \frac{2 x-1-\sqrt{1-\rho }}{2 x-1+\sqrt{1-\rho }}\\
-\frac{4 x^2}{(2 x-1) \left(\rho +4 x^2-4 x\right)}+\frac{2 x^2-1}{(\rho -1) (x-1) (2 x-1)}+\frac{8 \left(x^3-x^2\right)}{\left(\rho +4 x^2-4 x\right)^2}+\frac{3}{2 (x-1)},\;\;&x<0.\\
\end{array}\right.
\end{eqnarray}
In the limit $\rho\to 0$, it reduces to:
\begin{eqnarray}
\Delta  \tilde f_{q/q,z}^{(1)}(x, \rho)|_{\rho\to 0}  =\frac{\alpha_s C_F}{2\pi} \left\{ \begin{array}{ll} \frac{\left(x^2+1\right) \ln\frac{x-1}{x}+x-1}{x-1},\;\;&x>1\\\frac{2 \left(x^2+1\right) \ln \frac{\rho}{4}+4 x^2+1}{2 (x-1)},\;\;&0<x<1\\-\frac{\left(x^2+1\right) \ln\frac{x-1}{x}+x-1}{x-1},\;\;&x<0.\\\end{array}\right.\end{eqnarray}

The matching coefficient can then be extracted as
\begin{eqnarray}
\Delta C^{(1)}_{qq}(x,r, \frac{p_z}{\mu}, \frac{p_z}{p_z^R})&=& \bigg[\Delta  \tilde f_{q/q,z}^{(1)}(x, \rho\to 0)   -  \Delta f_{q/q}^{(1)}\bigg(x, \frac{\mu^2}{-p^2}\bigg)- (\Delta\tilde f_{q/q}^{(1)})_{C.T.}\bigg]_+,
\end{eqnarray}
where the counterterm in the RI/MOM scheme is determined as:
\begin{eqnarray}
(\Delta\tilde f_{q/q}^{(1)})_{C.T.}= \left|\frac{p_z}{p_z^R}\right|  \Delta  \tilde f_{q/q, z}^{(1)}\left(\frac{p_z}{p_z^R}(x-1)+1, r\right).
\end{eqnarray}

\subsection{Gluon in Quark}

The matrix element of the local gluon operator between the polarized quark state reads:
\begin{eqnarray}
&& \langle q(p)|F^{\mu\alpha}(0) F^{\nu\beta}(0)|q(p)\rangle =  -\frac{i\alpha_s C_F}{24\pi \epsilon}  \left({p}^{\alpha } \epsilon ^{\beta \mu \nu \delta }+{p}^{\beta } \epsilon ^{\alpha \mu \nu \delta }+{p}^{\mu } \epsilon ^{\alpha \beta \nu \delta }+{p}^{\nu } \epsilon ^{\alpha \beta \mu \delta }\right)
\bar u(p)  {\gamma}_{\delta }{\gamma}^5u(p),
\end{eqnarray}
where we have used the following identity
\begin{eqnarray}
\gamma^\mu\gamma^\nu\gamma^\alpha= \gamma^\mu g^{\alpha\nu} -\gamma^\nu g^{\alpha\mu}+\gamma^{\alpha}g^{\mu\nu} +i \epsilon^{\mu\nu\alpha\delta}\gamma_\delta\gamma^5.
\end{eqnarray}
Projecting onto $\Delta O_{g,R}^{(3)}(0,0)$ gives:
\begin{eqnarray}
&& \langle q(p)|\Delta O^{(3)}_{g,R}(0,0)|q(p)\rangle =  \frac{\alpha_s C_F}{3\pi\epsilon}  p^t p^z+{\cal O}(\epsilon^0),
\end{eqnarray}
which has the same momentum dependence as the quark matrix element of the quark quasi-PDF operator in Eq.~\eqref{eq:polarized_quasi_quark_pdf}.

The light-cone result for the polarized PDF is given as
 \begin{eqnarray}
x \Delta   f_{g/q}^{(1)}(x, \mu)
= \frac{\alpha_s C_F }{2\pi}    \left(x(x-2) \ln \frac{-p^2   (1-x) x}{\mu^2}+x^2-5x\right).
\end{eqnarray}
The corresponding quasi-PDF reads
\begin{eqnarray}
 x \Delta\tilde f_{g/q}^{(3,1)}(x, \rho)  &=&\frac{\alpha_s C_F}{2\pi} \left\{ \begin{array}{ll} \frac{(2 x-1) \left(\rho  (\rho +2)+4 (\rho +2) x^2+4 \left(\rho ^2-2 \rho -2\right) x\right)}{4 (\rho -1)^2 \left(\rho +4 x^2-4 x\right)}\\
 +\frac{\left(-(\rho -4) \rho +4 (\rho +2) x^2-4 \left(\rho ^2-2 \rho +4\right) x\right) \ln \frac{2 x-1-\sqrt{1-\rho }}{2 x-1+\sqrt{1-\rho }}}{8 (1-\rho )^{5/2}},\;\;&x>1\\
 \frac{\rho +12 x^2+4 (\rho -4) x+2}{4 (\rho -1)^2}+\frac{\ln \frac{1-\sqrt{1-\rho }}{1+\sqrt{1-\rho}} \left(-(\rho -4) \rho +4 (\rho +2) x^2-4 \left(\rho ^2-2 \rho +4\right) x\right)}{8 (1-\rho )^{5/2}},\;\;&0<x<1\\
 -\frac{(2 x-1) \left(\rho  (\rho +2)+4 (\rho +2) x^2+4 \left(\rho ^2-2 \rho -2\right) x\right)}{4 (\rho -1)^2 \left(\rho +4 x^2-4 x\right)}\\
 +\frac{\left((\rho -4) \rho -4 (\rho +2) x^2+4 \left(\rho ^2-2 \rho +4\right) x\right) \ln \frac{2 x-1-\sqrt{1-\rho }}{2 x-1+\sqrt{1-\rho }}}{8 (1-\rho )^{5/2}},\;\;&x<0.\\\end{array}\right.
\end{eqnarray}
 In the limit $\rho\to 0$, we have
 \begin{eqnarray}
  x\Delta \tilde f_{g/q}^{(3,1)}(x, \rho\to 0) &=\frac{\alpha_s C_F}{2\pi} \left\{ \begin{array}{ll} \frac{1}{2} \left(2 x+2 (x-2) x \ln\frac{x-1}{x}-1\right),\;\;&x>1\\\frac{1}{2} \left(2 (x-2) x \ln \frac{\rho}{4}+6 x^2-8 x+1\right),\;\;&0<x<1\\\frac{1}{2} \left(-2 x-2 (x-2) x \ln\frac{x-1}{x}+1\right),\;\;&x<0.\\\end{array}\right.\end{eqnarray}

 The matching coefficient can be extracted as
\begin{eqnarray}
x \Delta C_{g/q}^{(3,1)}(x,r, \frac{p_z}{\mu}, \frac{p_z}{p_z^R})&=& \bigg[x \Delta \tilde f_{g/q}^{(3,1)}(x, \rho\to 0)   - x  \Delta f_{g/q}^{(1)}\bigg(x, \frac{\mu^2}{-p^2}\bigg)- (x\Delta\tilde f_{g/q}^{(3,1)})_{C.T.}\bigg],
\end{eqnarray}
where  the counterterm in the RI/MOM scheme is determined as:
\begin{eqnarray}
(x\Delta\tilde f_{g/q}^{(3,1)})_{C.T.}= \left|\frac{p_z}{p_z^R}\right|  x\Delta\tilde f_{g/q}^{(3,1)}\left(\frac{p_z}{p_z^R}(x-1)+1, r\right).
\end{eqnarray}

\subsection{Quark in Gluon}

In this case,  the light-cone result is
\begin{eqnarray}
   \Delta f_{q/g}^{(1)}(x, \mu)
    = \frac{\alpha_s T_f}{2\pi}   \left((1-2 x) \ln \frac{-p^2   (1-x) x}{\mu^2}-4 x+1\right),
\end{eqnarray}
whereas the quasi PDF result reads
\begin{eqnarray}
\Delta   \tilde  f_{q/g}^{(1)}   =\frac{\alpha_sT_f}{2\pi} \left\{ \begin{array}{ll} -\frac{\rho +8 x^2+2 (\rho -4) x}{ 1-\rho  \left(\rho +4 x^2-4 x\right)}-\frac{\rho +4 x-2}{2(1-\rho)^{3/2}} \ln \frac{2 x-1-\sqrt{1-\rho }}{2 x-1+\sqrt{1-\rho }},\;\;&x>1\\
\frac{1-4 x}{  {1-\rho }}-\frac{\rho +4 x-2}{2(1-\rho)^{3/2}}\ln \frac{1-\sqrt{1-\rho }}{1+\sqrt{1-\rho }} ,\;\;&0<x<1\\
\frac{\rho +8 x^2+2 (\rho -4) x}{  {1-\rho } \left(\rho +4 x^2-4 x\right)}+\frac{\rho +4 x-2 }{2 (1-\rho)^{3/2}}\ln \frac{2 x-1-\sqrt{1-\rho }}{2 x-1+\sqrt{1-\rho }},\;\;&x<0.
\end{array}\right.
\end{eqnarray}
In the limit $\rho\to0$, we have
\begin{eqnarray}
\Delta\tilde f_{q/g}^{(1)}(x,\rho\to 0) =\frac{\alpha_sT_f}{2\pi} \left\{ \begin{array}{ll}    (1-2 x) \ln\frac{x-1}{x}-2 ,\;\;&x>1\\   (1-2 x) \ln \frac{\rho}{4}-4 x+1 ,\;\;&0<x<1\\\left(2x-1\right) \ln\frac{x-1}{x}+1,\;\;&x<0.\end{array}\right.
\end{eqnarray}

The matching coefficient is then given by
\begin{eqnarray}
\Delta C^{(1)}_{qg}(x,r, \frac{p_z}{\mu}, \frac{p_z}{p_z^R})&=& \bigg[\Delta \tilde f_{q/g}^{(1)}(x, \rho\to 0)   -   \Delta f_{q/g}^{(1)}\bigg(x, \frac{\mu^2}{-p^2}\bigg)- (\Delta\tilde f_{q/g}^{(1)})_{C.T.}\bigg],
\end{eqnarray}
with
\begin{eqnarray}
(\Delta \tilde f_{q/g}^{(1)})_{C.T.}= \left|\frac{p_z}{p_z^R}\right|  \Delta \tilde f_{q/g}^{(1)}\left(\frac{p_z}{p_z^R}(x-1)+1, r\right).
\end{eqnarray}

\section{Conclusion}
\label{conclusion}

In this paper, we have studied how to extract the flavor-singlet quark PDF and the gluon PDF from LaMET, both in the unpolarized and in the polarized case. After briefly reviewing the auxiliary ``heavy quark'' formalism used in our earlier work to prove the multiplicative renormalizability of quark and gluon quasi-PDF operators, we explained how a nonperturbative RI/MOM renormalization can be carried out for the quark and gluon quasi-PDFs on the lattice in the presence of mixing. Using OPE, we also derived the factorization formulas that connect them to the usual quark and gluon PDFs in $\rm \overline{MS}$ scheme. We then performed a one-loop calculation of the hard matching kernel appearing in the factorization. We found that certain gluon quasi-PDF operators are more favorable than others in the sense that the mixing with gauge variant operators can be avoided. We then focused on these operators and presented the corresponding one-loop matching kernel. Our results can be used to extract the flavor-singlet quark PDFs as well as the gluon PDFs from lattice simulations of the corresponding quasi-PDFs. We therefore completed the procedure of extracting quark and gluon PDFs from LaMET at leading power accuracy in the hadron momentum.

It is interesting to note that the matrix elements of those non-favorable gluon quasi-PDF operators have nontrivial momentum dependence in their asymptotic behavior at large $x$, which is also exhibited in the UV divergences of their local limit. This is a sign of the potential mixing with gauge variant operators. For these operators, it shall also be possible to work out an appropriate RI/MOM renormalization and matching, but one needs to take into account the gauge variant operators that are allowed to mix with the original operators. This makes the situation much more complicated and is beyond the scope of the present paper. We leave it to future work.

\section*{Acknowledgments}
We thank Vladimir Braun, Jiunn-Wei Chen, Xiangdong Ji, Yi-Zhuang Liu, Yu-Sheng Liu, Jian-Wei Qiu, Anatoly Radyushkin, Andreas Sch\"afer, Yi-Bo Yang, Feng Yuan, and  Yong Zhao for helpful discussions. This work is supported  in part  by  National  Natural
Science Foundation of China under Grant
 No.11575110,  11735010, 11705092, 11911530088, by Natural Science Foundation of Shanghai under Grant  No.~15DZ2272100, by Natural Science Foundation of Jiangsu under
Grant No.~BK20171471, and by the SFB/TRR-55 grant ``Hadron Physics from Lattice QCD". The work  of SZ is also supported by Jefferson Science Associates, LLC under  U.S. DOE Contract \#DE-AC05-06OR23177
and by U.S. DOE Grant \#DE-FG02-97ER41028.

\appendix
\label{appendix}

\section{One-loop Results in General $R_\xi$ Gauge }
\label{sec:oneloopRxi}

In this Appendix, we present the results for the one-loop matrix elements of all gluon quasi-PDF operators in general $R_\xi$ gauge. The matrix elements of the quark quasi-PDF operators do not depend on the choice of the gluon quasi-PDF operators and therefore will remain the same as those given the main text.
For the gluon matrix elements of the gluon quasi-PDF operators, the distribution part reads
  {\begin{align}x\tilde f_{g/g}^{(1,1)}(x,\rho) &=\frac{\alpha_s C_A}{2\pi} \left\{ \begin{array}{ll} -\frac{4 (x-1) x^2}{(2 x-1)^2 \left(\rho +4 x^2-4 x\right)}+\frac{2 \left(2 x^4-8 x^3+6 x^2-x\right)}{(\rho -1) (x-1) (2 x-1)^2}+\frac{12 x+1}{6 (\rho -1)^2}-\frac{(2 x-1)^2}{2 (\rho -1)^3}-\frac{1}{x-1},\;\;&x>1\nonumber\\-\frac{2 \left(2 x^2-x\right)}{(\rho -1) (x-1)}+\frac{8 x^3+12 x^2-6 x-1}{6 (\rho -1)^2}+\frac{-8 x^3+12 x^2-6 x+1}{2 (\rho -1)^3}+\frac{1}{x-1},\;\;&0<x<1\nonumber\\\frac{4 (x-1) x^2}{(2 x-1)^2 \left(\rho +4 x^2-4 x\right)}-\frac{2 \left(2 x^4-8 x^3+6 x^2-x\right)}{(\rho -1) (x-1) (2 x-1)^2}+\frac{-12 x-1}{6 (\rho -1)^2}+\frac{(2 x-1)^2}{2 (\rho -1)^3}+\frac{1}{x-1},\;\;&x<0\nonumber\\\end{array}\right.\nonumber\\&+\frac{\alpha_s C_A}{2\pi}  \left\{ \begin{array}{ll} \frac{\left(-6 \rho ^3+19 \rho ^2-20 \rho +8 x^4-4 \left(\rho ^2+4\right) x^3+6 \left(3 \rho ^2-4 \rho +4\right) x^2+\left(2 \rho ^3-17 \rho ^2+24 \rho -16\right) x+8\right) \ln \frac{2 x-1-\sqrt{1-\rho }}{2 x-1+\sqrt{1-\rho }}}{4 (1-\rho )^{7/2} (x-1)},\;\;&x>1\nonumber\\\frac{\ln \left(\frac{1-\sqrt{1-\rho }}{\sqrt{1-\rho}+1}\right) \left(-6 \rho ^3+19 \rho ^2-20 \rho +8 x^4-4 \left(\rho ^2+4\right) x^3+6 \left(3 \rho ^2-4 \rho +4\right) x^2+\left(2 \rho ^3-17 \rho ^2+24 \rho -16\right) x+8\right)}{4 (1-\rho )^{7/2} (x-1)},\;\;&0<x<1\nonumber\\-\frac{\left(-6 \rho ^3+19 \rho ^2-20 \rho +8 x^4-4 \left(\rho ^2+4\right) x^3+6 \left(3 \rho ^2-4 \rho +4\right) x^2+\left(2 \rho ^3-17 \rho ^2+24 \rho -16\right) x+8\right) \ln \frac{2 x-1-\sqrt{1-\rho }}{2 x-1+\sqrt{1-\rho }}}{4 (1-\rho )^{7/2} (x-1)},\;\;&x<0\nonumber\\\end{array}\right.\nonumber\\&+\frac{\alpha_s C_A}{2\pi}  \left\{ \begin{array}{ll} \frac{(\xi -1) \rho ^2 \left(\rho ^2+8 x^4-20 x^3+2 (2 \rho +7) x^2-(6 \rho +1) x\right)}{2 (\rho -1)^2 (x-1) \left(\rho +4 x^2-4 x\right)^2}+\frac{(\xi -1) \rho ^2 \ln \frac{2 x-1-\sqrt{1-\rho }}{2 x-1+\sqrt{1-\rho }}}{4 (1-\rho )^{5/2}},\;\;&x>1\\-\frac{(\xi -1) \left(\rho ^2-2 \rho  x+x\right)}{2 (\rho -1)^2 (x-1)}+\frac{(\xi -1) \rho ^2 \ln \left(\frac{1-\sqrt{1-\rho }}{\sqrt{1-\rho}+1}\right)}{4 (1-\rho )^{5/2}},\;\;&0<x<1\\-\frac{(\xi -1) \rho ^2 \left(\rho ^2+8 x^4-20 x^3+2 (2 \rho +7) x^2-(6 \rho +1) x\right)}{2 (\rho -1)^2 (x-1) \left(\rho +4 x^2-4 x\right)^2}-\frac{(\xi -1) \rho ^2 \ln \frac{2 x-1-\sqrt{1-\rho }}{2 x-1+\sqrt{1-\rho }}}{4 (1-\rho )^{5/2}},\;\;&x<0,\end{array}\right.
  \end{align}

  {\begin{align}x\tilde f_{g/g}^{(2,1)}(x,\rho) &=\frac{\alpha_s C_A}{2\pi} \left\{ \begin{array}{ll} -\frac{4 (x-1) x^2}{\rho +4 x^2-4 x}+\frac{-12 x^2-x+10}{6 (\rho -1) (x-1)}+\frac{(2 x-1)^2}{2 (\rho -1)^2}-\frac{1}{2 (x-1)}+\frac{(\xi -1) \rho ^2 x (2 x-1)}{2 (x-1) \left(\rho +4 x^2-4 x\right)^2},\;\;&x>1\nonumber\\\frac{4 x^2-2 x+1}{2 (x-1)}+\frac{8 x^3-12 x^2+6 x-1}{2 (\rho -1)^2}+\frac{-8 x^4-4 x^3-6 x^2+25 x-10}{6 (\rho -1) (x-1)}-\frac{(\xi -1) x}{2 (x-1)},\;\;&0<x<1\nonumber\\\frac{4 (x-1) x^2}{\rho +4 x^2-4 x}+\frac{12 x^2+x-10}{6 (\rho -1) (x-1)}-\frac{(2 x-1)^2}{2 (\rho -1)^2}+\frac{1}{2 (x-1)}-\frac{(\xi -1) \rho ^2 x (2 x-1)}{2 (x-1) \left(\rho +4 x^2-4 x\right)^2},\;\;&x<0\nonumber\\\end{array}\right.\nonumber\\
  &+\frac{\alpha_s C_A}{2\pi}  \left\{ \begin{array}{ll} -\frac{\left((\rho -2)^3-8 x^4+4 \left(\rho ^2+4\right) x^3-2 \left(5 \rho ^2-8 \rho +12\right) x^2+\left(3 \rho ^2-12 \rho +16\right) x\right) \ln \frac{2 x-1-\sqrt{1-\rho }}{2 x-1+\sqrt{1-\rho }}}{4 (1-\rho )^{5/2} (x-1)},\;\;&x>1\\-\frac{\ln \left(\frac{1-\sqrt{1-\rho }}{\sqrt{1-\rho}+1}\right) \left((\rho -2)^3-8 x^4+4 \left(\rho ^2+4\right) x^3-2 \left(5 \rho ^2-8 \rho +12\right) x^2+\left(3 \rho ^2-12 \rho +16\right) x\right)}{4 (1-\rho )^{5/2} (x-1)},\;\;&0<x<1\\\frac{\left((\rho -2)^3-8 x^4+4 \left(\rho ^2+4\right) x^3-2 \left(5 \rho ^2-8 \rho +12\right) x^2+\left(3 \rho ^2-12 \rho +16\right) x\right) \ln \frac{2 x-1-\sqrt{1-\rho }}{2 x-1+\sqrt{1-\rho }}}{4 (1-\rho )^{5/2} (x-1)},\;\;&x<0, \\\end{array}\right.\end{align}

{\begin{align}x\tilde f_{g/g}^{(3,1)}(x,\rho) &=\frac{\alpha_s C_A}{2\pi} \left\{ \begin{array}{ll} -\frac{4 \left(x^3-x^2\right)}{(2 x-1) \left(\rho +4 x^2-4 x\right)}+\frac{8 x^4-16 x^3-22 x^2+34 x-9}{4 (\rho -1) (x-1) (2 x-1)}+\frac{3 x (2 x-1)}{2 (\rho -1)^2}-\frac{2 x+1}{4 (x-1)}+\frac{(\xi -1) \rho ^2 x}{2 (x-1) \left(\rho +4 x^2-4 x\right)^2},\;\;&x>1\nonumber\\\frac{-30 x^2+34 x-9}{4 (\rho -1) (x-1)}+\frac{3 \left(4 x^3-4 x^2+x\right)}{2 (\rho -1)^2}+\frac{4 x+1}{4 (x-1)}-\frac{(\xi -1) x}{2 (x-1)},\;\;&0<x<1\nonumber\\\frac{4 \left(x^3-x^2\right)}{(2 x-1) \left(\rho +4 x^2-4 x\right)}+\frac{-8 x^4+16 x^3+22 x^2-34 x+9}{4 (\rho -1) (x-1) (2 x-1)}-\frac{3 x (2 x-1)}{2 (\rho -1)^2}+\frac{2 x+1}{4 (x-1)}-\frac{(\xi -1) \rho ^2 x}{2 (x-1) \left(\rho +4 x^2-4 x\right)^2},\;\;&x<0\nonumber\\\end{array}\right.\nonumber\\
&+\frac{\alpha_s C_A}{2\pi}  \left\{ \begin{array}{ll} \frac{\left(-(\rho -4)^2 (\rho -1)+8 (\rho +2) x^4-16 (\rho +2) x^3-2 \left(\rho ^2+8 \rho -24\right) x^2+\left(6 \rho ^2+20 \rho -32\right) x\right) \ln \frac{2 x-1-\sqrt{1-\rho }}{2 x-1+\sqrt{1-\rho }}}{8 (1-\rho )^{5/2} (x-1)},\;\;&x>1\\\frac{\ln \left(\frac{1-\sqrt{1-\rho }}{\sqrt{1-\rho}+1}\right) \left(-(\rho -4)^2 (\rho -1)+8 (\rho +2) x^4-16 (\rho +2) x^3-2 \left(\rho ^2+8 \rho -24\right) x^2+\left(6 \rho ^2+20 \rho -32\right) x\right)}{8 (1-\rho )^{5/2} (x-1)},\;\;&0<x<1\\\frac{\left((\rho -4)^2 (\rho -1)-8 (\rho +2) x^4+16 (\rho +2) x^3+2 \left(\rho ^2+8 \rho -24\right) x^2+\left(-6 \rho ^2-20 \rho +32\right) x\right) \ln \frac{2 x-1-\sqrt{1-\rho }}{2 x-1+\sqrt{1-\rho }}}{8 (1-\rho )^{5/2} (x-1)},\;\;&x<0, \end{array}\right.
\end{align}

  {\begin{align}x\tilde f_{g/g}^{(4,1)}(x,\rho)&=\frac{\alpha_s C_A}{2\pi} \left\{ \begin{array}{ll} \frac{4 x^2-4 x-1}{2 (x-1)}+\frac{-4 x^3+4 x^2-2 x+1}{2 (\rho -1) (x-1)}-\frac{4 \left(3 x^3-5 x^2+2 x\right)}{\rho +4 x^2-4 x}+\frac{(\xi -1) \rho ^2 x (2 x-1)}{2 (x-1) \left(\rho +4 x^2-4 x\right)^2},\;\;&x>1\nonumber\\\frac{8 x^2-6 x+1}{2 (x-1)}+\frac{-8 x^4+12 x^3-8 x^2+4 x-1}{2 (\rho -1) (x-1)}-\frac{(\xi -1) x}{2 (x-1)},\;\;&0<x<1\nonumber\\-\frac{4 x^2-4 x-1}{2 (x-1)}+\frac{4 x^3-4 x^2+2 x-1}{2 (\rho -1) (x-1)}+\frac{4 \left(3 x^3-5 x^2+2 x\right)}{\rho +4 x^2-4 x}-\frac{(\xi -1) \rho ^2 x (2 x-1)}{2 (x-1) \left(\rho +4 x^2-4 x\right)^2},\;\;&x<0\nonumber\\\end{array}\right.\nonumber\\
  &+\frac{\alpha_s C_A}{2\pi}  \left\{ \begin{array}{ll} \frac{\left(\rho ^2-8 \rho +8 x^4+4 (\rho -4) x^3-8 (2 \rho -3) x^2+4 (3 \rho -4) x+8\right) \ln \frac{2 x-1-\sqrt{1-\rho }}{2 x-1+\sqrt{1-\rho }}}{4 (1-\rho )^{3/2} (x-1)},\;\;&x>1\\\frac{\ln \left(\frac{1-\sqrt{1-\rho }}{\sqrt{1-\rho}+1}\right) \left(\rho ^2-8 \rho +8 x^4+4 (\rho -4) x^3-8 (2 \rho -3) x^2+4 (3 \rho -4) x+8\right)}{4 (1-\rho )^{3/2} (x-1)},\;\;&0<x<1\\-\frac{\left(\rho ^2-8 \rho +8 x^4+4 (\rho -4) x^3-8 (2 \rho -3) x^2+4 (3 \rho -4) x+8\right) \ln \frac{2 x-1-\sqrt{1-\rho }}{2 x-1+\sqrt{1-\rho }}}{4 (1-\rho )^{3/2} (x-1)},\;\;&x<0, \\\end{array}\right.
 \end{align}

    {\begin{align}x\Delta\tilde f_{g/g}^{(1,1)}(x,\rho) &=\frac{\alpha_s C_A}{2\pi} \left\{ \begin{array}{ll} -\frac{4 \left(x^3-x^2\right)}{(2 x-1)^2 \left(\rho +4 x^2-4 x\right)}+\frac{20 x^4-52 x^3+31 x^2-2 x-1}{2 (\rho -1) (x-1) (2 x-1)^2}+\frac{10 x-3}{2 (\rho -1)^2}-\frac{1}{x-1},\;\;&x>1\nonumber\\\frac{-3 x^2-2 x+1}{2 (\rho -1) (x-1)}+\frac{20 x^2-16 x+3}{2 (\rho -1)^2}+\frac{1}{x-1},\;\;&0<x<1\nonumber\\\frac{4 \left(x^3-x^2\right)}{(2 x-1)^2 \left(\rho +4 x^2-4 x\right)}+\frac{-20 x^4+52 x^3-31 x^2+2 x+1}{2 (\rho -1) (x-1) (2 x-1)^2}+\frac{3-10 x}{2 (\rho -1)^2}+\frac{1}{x-1},\;\;&x<0\nonumber\\\end{array}\right.\nonumber\\&+\frac{\alpha_s C_A}{2\pi}  \left\{ \begin{array}{ll} \frac{\left(\rho  (5 \rho -8)+4 (\rho +4) x^3-12 (\rho +2) x^2+\left(-\rho ^2+4 \rho +16\right) x\right) \ln \frac{2 x-1-\sqrt{1-\rho }}{2 x-1+\sqrt{1-\rho }}}{4 (1-\rho )^{5/2} (x-1)},\;\;&x>1\nonumber\\\frac{\ln \left(\frac{1-\sqrt{1-\rho }}{\sqrt{1-\rho}+1}\right) \left(\rho  (5 \rho -8)+4 (\rho +4) x^3-12 (\rho +2) x^2+\left(-\rho ^2+4 \rho +16\right) x\right)}{4 (1-\rho )^{5/2} (x-1)},\;\;&0<x<1\nonumber\\\frac{\left((8-5 \rho ) \rho -4 (\rho +4) x^3+12 (\rho +2) x^2+\left(\rho ^2-4 \rho -16\right) x\right) \ln \frac{2 x-1-\sqrt{1-\rho }}{2 x-1+\sqrt{1-\rho }}}{4 (1-\rho )^{5/2} (x-1)},\;\;&x<0\nonumber\\\end{array}\right.\nonumber\\&+\frac{\alpha_s C_A}{2\pi}  \left\{ \begin{array}{ll} \frac{(\xi -1) \rho ^2 \left(\rho ^2+8 x^4-20 x^3+2 (2 \rho +7) x^2-(6 \rho +1) x\right)}{2 (\rho -1)^2 (x-1) \left(\rho +4 x^2-4 x\right)^2}+ \frac{(\xi -1) \rho ^2 \ln \frac{2 x-1-\sqrt{1-\rho }}{2 x-1+\sqrt{1-\rho }}}{4 (1-\rho )^{5/2}},\;\;&x>1\\-\frac{(\xi -1) \left(\rho ^2-2 \rho  x+x\right)}{2 (\rho -1)^2 (x-1)}+\frac{(\xi -1) \rho ^2 \ln \left(\frac{1-\sqrt{1-\rho }}{\sqrt{1-\rho}+1}\right)}{4 (1-\rho )^{5/2}},\;\;&0<x<1\\-\frac{(\xi -1) \rho ^2 \left(\rho ^2+8 x^4-20 x^3+2 (2 \rho +7) x^2-(6 \rho +1) x\right)}{2 (\rho -1)^2 (x-1) \left(\rho +4 x^2-4 x\right)^2}-\frac{(\xi -1) \rho ^2 \ln \frac{2 x-1-\sqrt{1-\rho }}{2 x-1+\sqrt{1-\rho }}}{4 (1-\rho )^{5/2}},\;\;&x<0,\end{array}\right. \end{align}

{\begin{align}x\Delta\tilde f_{g/g}^{(2,1)}(x,\rho) &=\frac{\alpha_s C_A}{2\pi} \left\{ \begin{array}{ll} \frac{9 x-10 x^2}{2 (\rho -1) (x-1)}-\frac{4 \left(x^3-x^2\right)}{\rho +4 x^2-4 x}+\frac{(x-2) x}{2 (x-1)}+\frac{x \left((\rho -4)^2+4 (\rho +4) x^2-4 (\rho +6) x\right) \ln \frac{2 x-1-\sqrt{1-\rho }}{2 x-1+\sqrt{1-\rho }}}{4 (1-\rho )^{3/2} (x-1)},\;\;&x>1\nonumber\\\frac{3 x^2}{2 (x-1)}+\frac{-20 x^3+28 x^2-9 x}{2 (\rho -1) (x-1)}+\frac{x \ln \left(\frac{1-\sqrt{1-\rho }}{\sqrt{1-\rho}+1}\right) \left((\rho -4)^2+4 (\rho +4) x^2-4 (\rho +6) x\right)}{4 (1-\rho )^{3/2} (x-1)},\;\;&0<x<1\\\frac{10 x^2-9 x}{2 (\rho -1) (x-1)}+\frac{4 \left(x^3-x^2\right)}{\rho +4 x^2-4 x}-\frac{(x-2) x}{2 (x-1)}-\frac{x \left((\rho -4)^2+4 (\rho +4) x^2-4 (\rho +6) x\right) \ln \frac{2 x-1-\sqrt{1-\rho }}{2 x-1+\sqrt{1-\rho }}}{4 (1-\rho )^{3/2} (x-1)},\;\;&x<0\nonumber\\\end{array}\right.\\
&+\frac{\alpha_s C_A}{2\pi}  \left\{ \begin{array}{ll} \frac{(\xi -1) \rho ^2 x (2 x-1)}{2 (x-1) \left(\rho +4 x^2-4 x\right)^2},\;\;&x>1\\-\frac{(\xi -1) x}{2 (x-1)},\;\;&0<x<1\\-\frac{(\xi -1) \rho ^2 x (2 x-1)}{2 (x-1) \left(\rho +4 x^2-4 x\right)^2},\;\;&x<0,\end{array}\right.\end{align}

  {\begin{align}x\Delta\tilde f_{g/g}^{(3,1)}(x,\rho) &=\frac{\alpha_s C_A}{2\pi} \left\{ \begin{array}{ll} \frac{-8 x^3-8 x^2+14 x-3}{4 (\rho -1) (x-1) (2 x-1)}-\frac{4 \left(x^3-x^2\right)}{(2 x-1) \left(\rho +4 x^2-4 x\right)}+\frac{3 (2 x-1)}{2 (\rho -1)^2}-\frac{2 x+1}{4 (x-1)}+\frac{(\xi -1) \rho ^2 x}{2 (x-1) \left(\rho +4 x^2-4 x\right)^2},\;\;&x>1\nonumber\\\frac{3 \left(4 x^2-4 x+1\right)}{2 (\rho -1)^2}+\frac{-16 x^3+8 x^2+6 x-3}{4 (\rho -1) (x-1)}+\frac{4 x+1}{4 (x-1)}-\frac{(\xi -1) x}{2 (x-1)},\;\;&0<x<1\nonumber\\\frac{8 x^3+8 x^2-14 x+3}{4 (\rho -1) (x-1) (2 x-1)}+\frac{4 \left(x^3-x^2\right)}{(2 x-1) \left(\rho +4 x^2-4 x\right)}-\frac{3 (2 x-1)}{2 (\rho -1)^2}+\frac{2 x+1}{4 (x-1)}-\frac{(\xi -1) \rho ^2 x}{2 (x-1) \left(\rho +4 x^2-4 x\right)^2},\;\;&x<0\nonumber\\\end{array}\right.\nonumber\\
  &+\frac{\alpha_s C_A}{2\pi}  \left\{ \begin{array}{ll} -\frac{\left(\rho  \left(\rho ^2-3 \rho +8\right)+8 (\rho -4) x^3+8 \left(\rho ^2-\rho +6\right) x^2-2 \left(9 \rho ^2-10 \rho +16\right) x\right) \ln \frac{2 x-1-\sqrt{1-\rho }}{2 x-1+\sqrt{1-\rho }}}{8 (1-\rho )^{5/2} (x-1)},\;\;&x>1\\-\frac{\ln \left(\frac{1-\sqrt{1-\rho }}{\sqrt{1-\rho}+1}\right) \left(\rho  \left(\rho ^2-3 \rho +8\right)+8 (\rho -4) x^3+8 \left(\rho ^2-\rho +6\right) x^2-2 \left(9 \rho ^2-10 \rho +16\right) x\right)}{8 (1-\rho )^{5/2} (x-1)},\;\;&0<x<1\\\frac{\left(\rho  \left(\rho ^2-3 \rho +8\right)+8 (\rho -4) x^3+8 \left(\rho ^2-\rho +6\right) x^2-2 \left(9 \rho ^2-10 \rho +16\right) x\right) \ln \frac{2 x-1-\sqrt{1-\rho }}{2 x-1+\sqrt{1-\rho }}}{8 (1-\rho )^{5/2} (x-1)},\;\;&x<0. \end{array}\right.
  \end{align}



The quark matrix elements of the gluon quasi-PDF operators are given as follows:
  \begin{align}
  x\tilde f_{g/q,t}^{(1,1)}(x,\rho) &=\frac{\alpha_s C_F}{2\pi} \left\{ \begin{array}{ll} \frac{\left(-5 \rho ^2+16 \rho +4 (\rho +2) x^2-12 \rho  x-8\right) \ln \frac{2 x-1-\sqrt{1-\rho }}{2 x-1+\sqrt{1-\rho }}}{8 (1-\rho )^{5/2}}+\frac{-5 \rho +2 (\rho +2) x+2}{4 (\rho -1)^2},\;\;&x>1\\\frac{\ln \left(\frac{1-\sqrt{1-\rho }}{\sqrt{1-\rho}+1}\right) \left(-5 \rho ^2+16 \rho +4 (\rho +2) x^2-12 \rho  x-8\right)}{8 (1-\rho )^{5/2}}+\frac{5 \rho +12 x^2-4 (2 \rho +1) x-2}{4 (\rho -1)^2},\;\;&0<x<1\\\frac{\left(5 \rho ^2-16 \rho -4 (\rho +2) x^2+12 \rho  x+8\right) \ln \frac{2 x-1-\sqrt{1-\rho }}{2 x-1+\sqrt{1-\rho }}}{8 (1-\rho )^{5/2}}+\frac{5 \rho -2 (\rho +2) x-2}{4 (\rho -1)^2},\;\;&x<0, \\\end{array}\right.\end{align}

\begin{align}
x\tilde f_{g/q,z}^{(1,1)}(x,\rho) &=\frac{\alpha_s C_F}{2\pi} \left\{ \begin{array}{ll} \frac{-3 \rho ^2+8 (\rho +2) x^3-4 (\rho +2)^2 x^2+2 \rho  (2 \rho +7) x}{2 (\rho -1)^3 \left(\rho +4 x^2-4 x\right)}\\-\frac{\left(5 \rho ^2-6 \rho +4 (\rho +2) x^2+\left(-6 \rho ^2+2 \rho -8\right) x+4\right) \ln \frac{2 x-1-\sqrt{1-\rho }}{2 x-1+\sqrt{1-\rho }}}{4 (1-\rho )^{7/2}},\;\;&x>1\\\frac{3 (2 x-1) (2 x-\rho )}{2 (\rho -1)^3}-\frac{\ln \left(\frac{1-\sqrt{1-\rho }}{\sqrt{1-\rho}+1}\right) \left(5 \rho ^2-6 \rho +4 (\rho +2) x^2+\left(-6 \rho ^2+2 \rho -8\right) x+4\right)}{4 (1-\rho )^{7/2}},\;\;&0<x<1\\\frac{\left(5 \rho ^2-6 \rho +4 (\rho +2) x^2+\left(-6 \rho ^2+2 \rho -8\right) x+4\right) \ln \frac{2 x-1-\sqrt{1-\rho }}{2 x-1+\sqrt{1-\rho }}}{4 (1-\rho )^{7/2}}\\+\frac{3 \rho ^2-8 (\rho +2) x^3+4 (\rho +2)^2 x^2-2 \rho  (2 \rho +7) x}{2 (\rho -1)^3 \left(\rho +4 x^2-4 x\right)},\;\;&x<0,\end{array}\right.\end{align}

\begin{align}
x\tilde f_{g/q,p}^{(1,1)} &=\frac{\alpha_s C_F}{2\pi} \left\{ \begin{array}{ll} \frac{\rho  (\rho -2 x) \left(\rho +12 x^2+2 (\rho -7) x+2\right)}{2 (\rho -1)^3 \left(\rho +4 x^2-4 x\right)}-\frac{\rho  (\rho -2 x) (\rho +6 x-4) \ln \frac{2 x-1-\sqrt{1-\rho }}{2 x-1+\sqrt{1-\rho }}}{4 (1-\rho )^{7/2}},\;\;&x>1\\\frac{(\rho -2 x) (-\rho +(4 \rho +2) x-2)}{2 (\rho -1)^3}-\frac{\rho  \ln \left(\frac{1-\sqrt{1-\rho }}{\sqrt{1-\rho}+1}\right) (\rho -2 x) (\rho +6 x-4)}{4 (1-\rho )^{7/2}},\;\;&0<x<1\\\frac{\rho  (\rho -2 x) (\rho +6 x-4) \ln \frac{2 x-1-\sqrt{1-\rho }}{2 x-1+\sqrt{1-\rho }}}{4 (1-\rho )^{7/2}}-\frac{\rho  (\rho -2 x) \left(\rho +12 x^2+2 (\rho -7) x+2\right)}{2 (\rho -1)^3 \left(\rho +4 x^2-4 x\right)},\;\;&x<0,\end{array}\right.\end{align}

  \begin{align}
  x\tilde f_{g/q,t}^{(2,1)}(x,\rho) &=\frac{\alpha_s C_F}{2\pi} \left\{ \begin{array}{ll} \frac{\rho  (\rho +2)-8 (\rho +2) x^3-4 (\rho -10) x^2-2 \left(5 \rho ^2-8 \rho +12\right) x}{4 (\rho -1) \left(\rho +4 x^2-4 x\right)}\\
  -\frac{\left((\rho -4) \rho -4 (\rho +2) x^2-4 (\rho -4) x\right) \ln \frac{2 x-1-\sqrt{1-\rho }}{2 x-1+\sqrt{1-\rho }}}{8 (1-\rho )^{3/2}},\;\;&x>1\\\frac{\rho +12 x^2+8 \rho  x-20 x+2}{4-4 \rho }-\frac{\ln \left(\frac{1-\sqrt{1-\rho }}{\sqrt{1-\rho}+1}\right) \left((\rho -4) \rho -4 (\rho +2) x^2-4 (\rho -4) x\right)}{8 (1-\rho )^{3/2}},\;\;&0<x<1\\\frac{-\rho  (\rho +2)+8 (\rho +2) x^3+4 (\rho -10) x^2+2 \left(5 \rho ^2-8 \rho +12\right) x}{4 (\rho -1) \left(\rho +4 x^2-4 x\right)}\\-\frac{\left(-(\rho -4) \rho +4 (\rho +2) x^2+4 (\rho -4) x\right) \ln \frac{2 x-1-\sqrt{1-\rho }}{2 x-1+\sqrt{1-\rho }}}{8 (1-\rho )^{3/2}},\;\;&x<0,\end{array}\right.\end{align}

  \begin{align}
  x\tilde f_{g/q,z}^{(2,1)}(x,\rho) &=\frac{\alpha_s C_F}{2\pi} \left\{ \begin{array}{ll} \frac{-3 (\rho -2) \rho -8 (\rho +2) x^3+4 \left(\rho ^2-4 \rho +12\right) x^2+\left(-8 \rho ^2+22 \rho -32\right) x}{2 (\rho -1)^2 \left(\rho +4 x^2-4 x\right)} \\
   -\frac{\left(3 \rho ^2-8 \rho +4 (\rho +2) x^2+2 \left(\rho ^2+\rho -8\right) x+8\right) \ln \frac{2 x-1-\sqrt{1-\rho }}{2 x-1+\sqrt{1-\rho }}}{4 (1-\rho )^{5/2}},\;\;&x>1\\\frac{3 \rho -12 x^2-10 \rho  x+22 x-6}{2 (\rho -1)^2}-\frac{\ln \left(\frac{1-\sqrt{1-\rho }}{\sqrt{1-\rho}+1}\right) \left(3 \rho ^2-8 \rho +4 (\rho +2) x^2+2 \left(\rho ^2+\rho -8\right) x+8\right)}{4 (1-\rho )^{5/2}},\;\;&0<x<1\\\frac{\left(3 \rho ^2-8 \rho +4 (\rho +2) x^2+2 \left(\rho ^2+\rho -8\right) x+8\right) \ln \frac{2 x-1-\sqrt{1-\rho }}{2 x-1+\sqrt{1-\rho }}}{4 (1-\rho )^{5/2}}\\
   +\frac{3 (\rho -2) \rho +8 (\rho +2) x^3-4 \left(\rho ^2-4 \rho +12\right) x^2+\left(8 \rho ^2-22 \rho +32\right) x}{2 (\rho -1)^2 \left(\rho +4 x^2-4 x\right)},\;\;&x<0,\end{array}\right.\end{align}

   \begin{align}
   x\tilde f_{g/q,p}^{(2,1)}(x,\rho) &=\frac{\alpha_s C_F}{2\pi} \left\{ \begin{array}{ll} \frac{\rho  \left(\rho ^2-2 \rho +12 x^2+4 (\rho -4) x+4\right) \ln \frac{2 x-1-\sqrt{1-\rho }}{2 x-1+\sqrt{1-\rho }}}{4 (1-\rho )^{5/2}}\\+\frac{\rho  \left((\rho -4) \rho +24 x^3+(8 \rho -44) x^2+2 \left(\rho ^2-2 \rho +10\right) x\right)}{2 (\rho -1)^2 \left(\rho +4 x^2-4 x\right)},\;\;&x>1\\\frac{-(\rho -4) \rho +(8 \rho +4) x^2+4 \left(\rho ^2-3 \rho -1\right) x}{2 (\rho -1)^2}+\frac{\rho  \ln \left(\frac{1-\sqrt{1-\rho }}{\sqrt{1-\rho}+1}\right) \left(\rho ^2-2 \rho +12 x^2+4 (\rho -4) x+4\right)}{4 (1-\rho )^{5/2}},\;\;&0<x<1\\-\frac{\rho  \left(\rho ^2-2 \rho +12 x^2+4 (\rho -4) x+4\right) \ln \frac{2 x-1-\sqrt{1-\rho }}{2 x-1+\sqrt{1-\rho }}}{4 (1-\rho )^{5/2}}\\-\frac{\rho  \left((\rho -4) \rho +24 x^3+(8 \rho -44) x^2+2 \left(\rho ^2-2 \rho +10\right) x\right)}{2 (\rho -1)^2 \left(\rho +4 x^2-4 x\right)},\;\;&x<0,\end{array}\right.\end{align}

  \begin{align}
  x\tilde f_{g/q,t}^{(3,1)}(x,\rho) &=\frac{\alpha_s C_F}{2\pi} \left\{ \begin{array}{ll} \frac{\left(3 \rho +4 x^2-4 x-2\right) \ln \frac{2 x-1-\sqrt{1-\rho }}{2 x-1+\sqrt{1-\rho }}}{4 (1-\rho )^{3/2}}+\frac{\rho -8 x^3+4 (\rho +2) x^2-6 \rho  x}{2 (\rho -1) \left(\rho +4 x^2-4 x\right)},\;\;&x>1\\\frac{\ln \left(\frac{1-\sqrt{1-\rho }}{\sqrt{1-\rho}+1}\right) \left(3 \rho +4 x^2-4 x-2\right)}{4 (1-\rho )^{3/2}}-\frac{(1-2 x)^2}{2 (\rho -1)},\;\;&0<x<1\\-\frac{\left(3 \rho +4 x^2-4 x-2\right) \ln \frac{2 x-1-\sqrt{1-\rho }}{2 x-1+\sqrt{1-\rho }}}{4 (1-\rho )^{3/2}}-\frac{\rho -8 x^3+4 (\rho +2) x^2-6 \rho  x}{2 (\rho -1) \left(\rho +4 x^2-4 x\right)},\;\;&x<0,\end{array}\right.\end{align}

  \begin{align}
  x\tilde f_{g/q,z}^{(3,1)}(x,\rho) &=\frac{\alpha_s C_F}{2\pi} \left\{ \begin{array}{ll} \frac{(5-2 \rho ) \rho -8 (\rho +2) x^3+4 (\rho +8) x^2-2 (\rho +8) x}{2 (\rho -1)^2 \left(\rho +4 x^2-4 x\right)}\\
  -\frac{\left(2 \rho ^2-5 \rho +4 (\rho +2) x^2-12 x+6\right) \ln \frac{2 x-1-\sqrt{1-\rho }}{2 x-1+\sqrt{1-\rho }}}{4 (1-\rho )^{5/2}},\;\;&x>1\\-\frac{\ln \left(\frac{1-\sqrt{1-\rho }}{\sqrt{1-\rho}+1}\right) \left(2 \rho ^2-5 \rho +4 (\rho +2) x^2-12 x+6\right)}{4 (1-\rho )^{5/2}}-\frac{(2 x-1) (2 \rho +6 x-5)}{2 (\rho -1)^2},\;\;&0<x<1\\\frac{\left(2 \rho ^2-5 \rho +4 (\rho +2) x^2-12 x+6\right) \ln \frac{2 x-1-\sqrt{1-\rho }}{2 x-1+\sqrt{1-\rho }}}{4 (1-\rho )^{5/2}}\\+\frac{\rho  (2 \rho -5)+8 (\rho +2) x^3-4 (\rho +8) x^2+2 (\rho +8) x}{2 (\rho -1)^2 \left(\rho +4 x^2-4 x\right)},\;\;&x<0,\end{array}\right.\end{align}

  \begin{align}
  x\tilde f_{g/q,p}^{(3,1)} &=\frac{\alpha_s C_F}{2\pi} \left\{ \begin{array}{ll} \frac{\rho  \left(\rho +12 x^2-12 x+2\right) \ln \frac{2 x-1-\sqrt{1-\rho }}{2 x-1+\sqrt{1-\rho }}}{4 (1-\rho )^{5/2}}+\frac{3 \rho  (2 x-1)}{2 (\rho -1)^2},\;\;&x>1\\\frac{3 \rho +(8 \rho +4) x^2-4 (2 \rho +1) x}{2 (\rho -1)^2}+\frac{\rho  \ln \left(\frac{1-\sqrt{1-\rho }}{\sqrt{1-\rho}+1}\right) \left(\rho +12 x^2-12 x+2\right)}{4 (1-\rho )^{5/2}},\;\;&0<x<1\\\frac{\rho  (3-6 x)}{2 (\rho -1)^2}-\frac{\rho  \left(\rho +12 x^2-12 x+2\right) \ln \frac{2 x-1-\sqrt{1-\rho }}{2 x-1+\sqrt{1-\rho }}}{4 (1-\rho )^{5/2}},\;\;&x<0,\end{array}\right.
  \end{align}

  \begin{align}
  x\tilde f_{g/q,t}^{(4,1)}(x,\rho) =0,\end{align}

  \begin{align}
  x\tilde f_{g/q,z}^{(4,1)}(x,\rho) &=\frac{\alpha_s C_F}{2\pi} \left\{ \begin{array}{ll} \frac{-\rho +4 x^3+2 (3 \rho -8) x^2-6 (\rho -2) x}{(\rho -1) \left(\rho +4 x^2-4 x\right)}-\frac{\left(-3 \rho +2 x^2-4 x+4\right) \ln \frac{2 x-1-\sqrt{1-\rho }}{2 x-1+\sqrt{1-\rho }}}{2 (1-\rho )^{3/2}},\;\;&x>1\\\frac{2 x^2-4 x+1}{\rho -1}-\frac{\ln \left(\frac{1-\sqrt{1-\rho }}{\sqrt{1-\rho}+1}\right) \left(-3 \rho +2 x^2-4 x+4\right)}{2 (1-\rho )^{3/2}},\;\;&0<x<1\\\frac{\rho -4 x^3+(16-6 \rho ) x^2+6 (\rho -2) x}{(\rho -1) \left(\rho +4 x^2-4 x\right)}-\frac{\left(3 \rho -2 x^2+4 x-4\right) \ln \frac{2 x-1-\sqrt{1-\rho }}{2 x-1+\sqrt{1-\rho }}}{2 (1-\rho )^{3/2}},\;\;&x<0,\end{array}\right.\end{align}
  \begin{align}x\tilde f_{g/q,p}^{(4,1)}(x,\rho) &=\frac{\alpha_s C_F}{2\pi} \left\{ \begin{array}{ll} \frac{\rho  \left(\rho -4 x^3+10 x^2-6 x\right)}{(\rho -1) \left(\rho +4 x^2-4 x\right)}-\frac{\rho  \left(\rho -2 x^2+4 x-2\right) \ln \frac{2 x-1-\sqrt{1-\rho }}{2 x-1+\sqrt{1-\rho }}}{2 (1-\rho )^{3/2}},\;\;&x>1\\-\frac{\rho +2 x^2-2 (\rho +1) x}{\rho -1}-\frac{\rho  \ln \left(\frac{1-\sqrt{1-\rho }}{\sqrt{1-\rho}+1}\right) \left(\rho -2 x^2+4 x-2\right)}{2 (1-\rho )^{3/2}},\;\;&0<x<1\\\frac{\rho  \left(\rho -2 x^2+4 x-2\right) \ln \frac{2 x-1-\sqrt{1-\rho }}{2 x-1+\sqrt{1-\rho }}}{2 (1-\rho )^{3/2}}-\frac{\rho  \left(\rho -4 x^3+10 x^2-6 x\right)}{(\rho -1) \left(\rho +4 x^2-4 x\right)},\;\;&x<0,\end{array}\right.\end{align}

  \begin{align}
x  \Delta\tilde f_{g/q,t}^{(1,1)}(x,\rho) &=\frac{\alpha_s C_F}{2\pi} \left\{ \begin{array}{ll} \frac{1-x}{(\rho -1)^2}-\frac{(x-1) (\rho +2 x-2) \ln \left(\frac{-\sqrt{1-\rho }+2 x-1}{\sqrt{1-\rho }+2 x-1}\right)}{2 (1-\rho )^{5/2}},\;\;&x>1\\-\frac{(x-1) (2 x-1)}{(\rho -1)^2}-\frac{(x-1) \ln \left(\frac{1-\sqrt{1-\rho }}{\sqrt{1-\rho}+1}\right) (\rho +2 x-2)}{2 (1-\rho )^{5/2}},\;\;&0<x<1\\\frac{x-1}{(\rho -1)^2}+\frac{(x-1) (\rho +2 x-2) \ln \left(\frac{-\sqrt{1-\rho }+2 x-1}{\sqrt{1-\rho }+2 x-1}\right)}{2 (1-\rho )^{5/2}},\;\;&x<0,\end{array}\right.\end{align}

\begin{align}
x\Delta\tilde f_{g/q,z}^{(1,1)}(x,\rho) &=\frac{\alpha_s C_F}{2\pi} \left\{ \begin{array}{ll} \frac{\left(\rho +2 (\rho +2) x^2-(\rho +8) x+2\right) \ln \left(\frac{-\sqrt{1-\rho }+2 x-1}{\sqrt{1-\rho }+2 x-1}\right)}{2 (1-\rho )^{5/2}}\\+\frac{-3 \rho +4 (\rho +2) x^3-4 (\rho +5) x^2+\left(2 \rho ^2+\rho +12\right) x}{(\rho -1)^2 \left(\rho +4 x^2-4 x\right)},\;\;&x>1\\\frac{6 x^2+(2 \rho -11) x+3}{(\rho -1)^2}+\frac{\ln \left(\frac{1-\sqrt{1-\rho }}{\sqrt{1-\rho}+1}\right) \left(\rho +2 (\rho +2) x^2-(\rho +8) x+2\right)}{2 (1-\rho )^{5/2}},\;\;&0<x<1\\\frac{3 \rho -4 (\rho +2) x^3+4 (\rho +5) x^2-\left(2 \rho ^2+\rho +12\right) x}{(\rho -1)^2 \left(\rho +4 x^2-4 x\right)}\\-\frac{\left(\rho +2 (\rho +2) x^2-(\rho +8) x+2\right) \ln \left(\frac{-\sqrt{1-\rho }+2 x-1}{\sqrt{1-\rho }+2 x-1}\right)}{2 (1-\rho )^{5/2}},\;\;&x<0,\end{array}\right.
\end{align}

  \begin{align}
  x\Delta\tilde f_{g/q,p}^{(1,1)}(x,\rho) &=\frac{\alpha_s C_F}{2\pi} \left\{ \begin{array}{ll} \frac{\rho  (x-1) \left(-3 \rho -12 x^2+2 (\rho +5) x\right)}{(1-\rho )^{5/2} \left(\rho +4 x^2-4 x\right)}+\frac{\rho  (x-1) (-\rho +6 x-2) \ln \frac{2 x-1-\sqrt{1-\rho }}{2 x-1+\sqrt{1-\rho }}}{2 (\rho -1)^3},\;\;&x>1\\\frac{\rho  (x-1) \ln \left(\frac{1-\sqrt{1-\rho }}{\sqrt{1-\rho}+1}\right) (-\rho +6 x-2)}{2 (\rho -1)^3}-\frac{(x-1) ((4 \rho +2) x-3 \rho )}{(1-\rho )^{5/2}},\;\;&0<x<1\\\frac{\rho  (x-1) \left(3 \rho +12 x^2-2 (\rho +5) x\right)}{(1-\rho )^{5/2} \left(\rho +4 x^2-4 x\right)}-\frac{\rho  (x-1) (-\rho +6 x-2) \ln \frac{2 x-1-\sqrt{1-\rho }}{2 x-1+\sqrt{1-\rho }}}{2 (\rho -1)^3},\;\;&x<0,\end{array}\right.\end{align}

\begin{align}
x\Delta\tilde f_{g/q,t}^{(2,1)}(x,\rho) &=\frac{\alpha_s C_F}{2\pi} \left\{ \begin{array}{ll} \frac{x \left(-3 \rho +4 x^2+4 (\rho -2) x+4\right)}{(\rho -1) \left(\rho +4 x^2-4 x\right)}-\frac{x (2 x-\rho ) \ln \left(\frac{-\sqrt{1-\rho }+2 x-1}{\sqrt{1-\rho }+2 x-1}\right)}{2 (1-\rho )^{3/2}},\;\;&x>1\\\frac{x (2 x-1)}{\rho -1}-\frac{x \ln \left(\frac{1-\sqrt{1-\rho }}{\sqrt{1-\rho}+1}\right) (2 x-\rho )}{2 (1-\rho )^{3/2}},\;\;&0<x<1\\-\frac{x \left(-3 \rho +4 x^2+4 (\rho -2) x+4\right)}{(\rho -1) \left(\rho +4 x^2-4 x\right)}-\frac{x (\rho -2 x) \ln \left(\frac{-\sqrt{1-\rho }+2 x-1}{\sqrt{1-\rho }+2 x-1}\right)}{2 (1-\rho )^{3/2}},\;\;&x<0,\end{array}\right.\end{align}
\begin{align}
x\Delta\tilde f_{g/q,z}^{(2,1)}(x,\rho) &=\frac{\alpha_s C_F}{2\pi} \left\{ \begin{array}{ll} \frac{x (\rho +2 (\rho +2) x-4) \ln \left(\frac{-\sqrt{1-\rho }+2 x-1}{\sqrt{1-\rho }+2 x-1}\right)}{2 (1-\rho )^{3/2}}-\frac{x \left(2 \rho ^2-3 \rho +4 (\rho +2) x^2-12 x+4\right)}{(\rho -1) \left(\rho +4 x^2-4 x\right)},\;\;&x>1\\\frac{x \ln \left(\frac{1-\sqrt{1-\rho }}{\sqrt{1-\rho}+1}\right) (\rho +2 (\rho +2) x-4)}{2 (1-\rho )^{3/2}}-\frac{x (2 \rho +6 x-5)}{\rho -1},\;\;&0<x<1\\\frac{x \left(2 \rho ^2-3 \rho +4 (\rho +2) x^2-12 x+4\right)}{(\rho -1) \left(\rho +4 x^2-4 x\right)}-\frac{x (\rho +2 (\rho +2) x-4) \ln \left(\frac{-\sqrt{1-\rho }+2 x-1}{\sqrt{1-\rho }+2 x-1}\right)}{2 (1-\rho )^{3/2}},\;\;&x<0,\end{array}\right.\end{align}

  \begin{align}x\Delta\tilde f_{g/q,p}^{(2,1)}(x,\rho) &=\frac{\alpha_s C_F}{2\pi} \left\{ \begin{array}{ll} -\frac{\rho  x \left(\rho +12 x^2+2 (\rho -7) x+2\right)}{(1-\rho )^{3/2} \left(\rho +4 x^2-4 x\right)}-\frac{\rho  x (\rho +6 x-4) \ln \frac{2 x-1-\sqrt{1-\rho }}{2 x-1+\sqrt{1-\rho }}}{2 (\rho -1)^2},\;\;&x>1\\-\frac{x (-\rho +(4 \rho +2) x-2)}{(1-\rho )^{3/2}}-\frac{\rho  x \ln \left(\frac{1-\sqrt{1-\rho }}{\sqrt{1-\rho}+1}\right) (\rho +6 x-4)}{2 (\rho -1)^2},\;\;&0<x<1\\\frac{\rho  x \left(\rho +12 x^2+2 (\rho -7) x+2\right)}{(1-\rho )^{3/2} \left(\rho +4 x^2-4 x\right)}+\frac{\rho  x (\rho +6 x-4) \ln \frac{2 x-1-\sqrt{1-\rho }}{2 x-1+\sqrt{1-\rho }}}{2 (\rho -1)^2},\;\;&x<0,\end{array}\right.\end{align}

  \begin{align}
  x\Delta\tilde f_{g/q,t}^{(3,1)}(x,\rho) &=\frac{\alpha_s C_F}{2\pi} \left\{ \begin{array}{ll} -\frac{\left(\rho ^2-2 \rho +4 (\rho +2) x^2-4 (\rho +2) x+4\right) \ln \left(\frac{-\sqrt{1-\rho }+2 x-1}{\sqrt{1-\rho }+2 x-1}\right)}{8 (1-\rho )^{5/2}}\\-\frac{(2 x-1) \left(-(\rho -4) \rho +4 (\rho +2) x^2-4 (\rho +2) x\right)}{4 (\rho -1)^2 \left(\rho +4 x^2-4 x\right)},\;\;&x>1\\
  \frac{\rho -12 x^2+12 x-4}{4 (\rho -1)^2}-\frac{\ln \left(\frac{1-\sqrt{1-\rho }}{\sqrt{1-\rho}+1}\right) \left(\rho ^2-2 \rho +4 (\rho +2) x^2-4 (\rho +2) x+4\right)}{8 (1-\rho )^{5/2}},\;\;&0<x<1\\
  \frac{\left(\rho ^2-2 \rho +4 (\rho +2) x^2-4 (\rho +2) x+4\right) \ln \left(\frac{-\sqrt{1-\rho }+2 x-1}{\sqrt{1-\rho }+2 x-1}\right)}{8 (1-\rho )^{5/2}}\\+\frac{(2 x-1) \left(-(\rho -4) \rho +4 (\rho +2) x^2-4 (\rho +2) x\right)}{4 (\rho -1)^2 \left(\rho +4 x^2-4 x\right)},\;\;&x<0,
  \end{array}\right.\end{align}

  \begin{align}
  x\Delta\tilde f_{g/q,z}^{(3,1)}(x,\rho) &=\frac{\alpha_s C_F}{2\pi} \left\{ \begin{array}{ll} \frac{(2 x-1) \left(3 \rho +4 (\rho +2) x^2+2 \left(\rho ^2-3 \rho -4\right) x\right)}{2 (\rho -1)^2 \left(\rho +4 x^2-4 x\right)}\\
  +\frac{\left(\rho +4 (\rho +2) x^2-2 \left(\rho ^2-\rho +6\right) x+2\right) \ln \left(\frac{-\sqrt{1-\rho }+2 x-1}{\sqrt{1-\rho }+2 x-1}\right)}{4 (1-\rho )^{5/2}},\;\;&x>1\\\frac{\ln \left(\frac{1-\sqrt{1-\rho }}{\sqrt{1-\rho}+1}\right) \left(\rho +4 (\rho +2) x^2-2 \left(\rho ^2-\rho +6\right) x+2\right)}{4 (1-\rho )^{5/2}}+\frac{12 x^2+2 (\rho -7) x+3}{2 (\rho -1)^2},\;\;&0<x<1\\-\frac{(2 x-1) \left(3 \rho +4 (\rho +2) x^2+2 \left(\rho ^2-3 \rho -4\right) x\right)}{2 (\rho -1)^2 \left(\rho +4 x^2-4 x\right)}\\
  -\frac{\left(\rho +4 (\rho +2) x^2-2 \left(\rho ^2-\rho +6\right) x+2\right) \ln \left(\frac{-\sqrt{1-\rho }+2 x-1}{\sqrt{1-\rho }+2 x-1}\right)}{4 (1-\rho )^{5/2}},\;\;&x<0,\end{array}\right.\end{align}

  \begin{align}
  x\Delta\tilde f_{g/q,p}^{(3,1)}(x,\rho) &=\frac{\alpha_s C_F}{2\pi} \left\{ \begin{array}{ll} \frac{\rho  \left(\rho +12 x^2-12 x+2\right) \ln \frac{2 x-1-\sqrt{1-\rho }}{2 x-1+\sqrt{1-\rho }}}{4 (\rho -1)^3}+\frac{\rho  (3-6 x)}{2 (1-\rho )^{5/2}},\;\;&x>1\\\frac{\rho  \ln\frac{1-\sqrt{1-\rho}}{1+\sqrt{1-\rho}} \left(\rho +12 x^2-12 x+2\right)}{4 (\rho -1)^3}-\frac{3 \rho +(8 \rho +4) x^2-4 (2 \rho +1) x}{2 (1-\rho )^{5/2}},\;\;&0<x<1\\
  \frac{3 \rho  (2 x-1)}{2 (1-\rho )^{5/2}}-\frac{\rho  \left(\rho +12 x^2-12 x+2\right) \ln \frac{2 x-1-\sqrt{1-\rho }}{2 x-1+\sqrt{1-\rho }}}{4 (\rho -1)^3},\;\;&x<0.\end{array}\right.
\end{align}

\end{document}